\documentclass[
 preprint,
amsmath,amssymb,superscriptaddress
]{revtex4-1}

\usepackage{graphicx}
\usepackage{lineno}
\usepackage{dcolumn}
\usepackage{bm}
\usepackage[normalem]{ulem}
\usepackage{xcolor}

\begin{document}

\title{Competing ferromagnetic and antiferromagnetic phases on the frustrated Ising honeycomb lattice}

\author{P.F. Dias}
\affiliation{Departamento de Física, Universidade Federal de Santa Maria, Santa Maria, 97105-900, RS, Brazil}


\author{F.M. Zimmer}
\affiliation{Instituto de Física, Universidade Federal de Mato Grosso do Sul, Campo Grande, 79070-900, MS, Brazil}

\author{N. G. Fytas}
\affiliation{School of Mathematics, Statistics and Actuarial Science, University of Essex, Colchester, CO4 3SQ, United Kingdom}

\author{M. Schmidt}
\affiliation{Departamento de Física, Universidade Federal de Santa Maria, Santa Maria, 97105-900, RS, Brazil}

\begin{abstract}
We investigate the frustrated $J_1$--$J_2$--$J_3$ Ising model on the honeycomb lattice, featuring first- and second-neighbor ferromagnetic couplings ($J_1>0$ and $J_2>0$) and third-neighbor antiferromagnetic interactions ($J_3<0$). Using the cluster mean-field method, we analyze the phase transitions in the regime $1/2 < J_2/J_1 \le 1$, where ferromagnetic and antiferromagnetic phases compete. Our results reveal that near the strongly frustrated limit $J_3/J_1 = -1$, the system exhibits order-by-disorder state selection, tricritical and bicritical behavior, critical endpoints, and two successive phase transitions. The ferromagnetic--paramagnetic transition remains  second order across the entire interaction range, whereas the antiferromagnetic--paramagnetic boundary shows a richer behavior, including both first- and second-order transitions as well as tricriticality. Increasing the second-neighbor coupling $J_2/J_1$ narrows the range of $J_3/J_1$ where first-order antiferromagnetic--paramagnetic transitions occur; beyond a certain threshold, only  second-order order--disorder transitions persist. Consequently, the tricritical point shifts toward $J_3/J_1 \approx -1$ as $J_2/J_1$ increases, culminating in a bicritical point where the antiferromagnetic, ferromagnetic, and paramagnetic phases meet.

\end{abstract} 
\maketitle

\section{Introduction}
\label{sec:Intro}

Low-dimensional magnetic systems have long attracted considerable attention due to their relevance to the fundamental physics of phase transitions and their potential technological applications~\cite{zhang20242d,Review2022-2DMaterials}. Significant progress in this field has been driven by the synthesis and characterization of van der Waals materials, many of which exhibit layered honeycomb-lattice magnetic structures~\cite{Review2022-VanDerWaals}. Notably, the magnetic long-range order in these compounds has been observed to persist down to the monolayer limit~\cite{FePS3-2016}, which, in light of the Mermin--Wagner theorem~\cite{PhysRevLett.17.1133}, implies the presence of spin anisotropy. Recent studies of van der Waals magnets such as FePS$_3$~\cite{FePS3-2016}, VBr$_3$~\cite{VBr3-2023}, CrI$_3$~\cite{huang2017layer}, and FePSe$_3$~\cite{chen2024thermal} provide compelling evidence of strong Ising anisotropy. Moreover, it has been suggested that competing exchange interactions extending up to third neighbors play a crucial role in determining the magnetic behavior of these systems~\cite{Review2022-VanDerWaals,chen2024thermal,WildesPRB-2020}. These experimental findings highlight the importance of magnetic frustration in honeycomb lattices and motivate theoretical investigations of the frustrated honeycomb Ising model.

In recent years, considerable theoretical effort has been devoted to understanding the phase transitions in the Ising model on the honeycomb lattice, with particular emphasis on the case involving first- ($J_1$) and second-neighbor ($J_2$) interactions. When ferromagnetic couplings between first neighbors ($J_1>0$) are considered, the model exhibits a ferromagnetic ground state for $g_2 \equiv J_2/J_1 > -1/4$. Early effective-field-theory studies suggested that thermal fluctuations could induce both first- and second-order phase transitions between the ferromagnetic (FE) and paramagnetic (PM) phases, implying the existence of a tricritical point~\cite{BOBAK20162693}. However, more recent results obtained through cluster mean-field (CMF) calculations~\cite{SCHMIDT2021168151}, correlated cluster mean-field~\cite{batista2025correlatedclustermeanfieldapproach}, partition-function-zero analysis~\cite{Entropy_2024}, and Monte Carlo simulations~\cite{Zukovic-2021,gessert2025,AZHARI2025108412,li2025} support a scenario in which only second-order FE--PM transitions occur in the $J_1$--$J_2$ Ising model. Monte Carlo simulations based on the Wang--Landau algorithm further indicate the emergence of a highly degenerate state for $g_2 < -1/4$~\cite{AZHARI2025108412}. It has been proposed that tricriticality may arise at the boundary between this degenerate state and the high-temperature paramagnetic phase~\cite{AZHARI2025108412}, although this picture has not been confirmed by more recent Metropolis-based simulations~\cite{li2025}. When third-neighbor interactions ($J_3$) are included, the system displays several distinct long-range ordered states depending on the relative strengths of $g_2$ and $g_3 \equiv J_3/J_1$. Recent CMF and Monte Carlo studies focusing on the regime $g_2 < 1/2$ have revealed a range of fluctuation-driven phenomena, including order-by-disorder state selection, anomalous specific-heat behavior, and tricriticality~\cite{Pietro2023-PRB,Pietro2024-MC}. Although restricted to specific parameter regimes, these works have provided valuable benchmarks for the theoretical description of honeycomb-lattice materials with strong easy-axis anisotropy.

In the present work, we aim to deepen the understanding of the $J_1$--$J_2$--$J_3$ Ising model on the honeycomb lattice by exploring the phase transitions that occur for $g_2 > 1/2$ (with $J_1 > 0$)—a parameter range largely overlooked in previous investigations. In this regime, both first- and second-neighbor interactions are ferromagnetic, while antiferromagnetic third-neighbor couplings introduce frustration, giving rise to competition between ferromagnetic and antiferromagnetic (AF) phases. Our analysis is carried out within the CMF framework, which has been successfully employed in several recent studies of frustrated magnetic systems~\cite{PhysRevLett.114.027201,Roos2024,PhysRevLett.125.057204,Rossato2023,PhysRevB.101.064403,PhysRevE.109.014144}. In particular, the CMF description of the FE--PM phase transition in the $J_1$--$J_2$ Ising model is consistent with the results of recent Monte Carlo simulations~\cite{SCHMIDT2021168151,Zukovic-2021,gessert2025,AZHARI2025108412,li2025}. Moreover, comparative studies of the CMF method and Monte Carlo simulations for the frustrated honeycomb Ising model with ferromagnetic $J_3$ interactions have shown that both approaches yield similar thermodynamic behavior~\cite{Pietro2024-MC}. These findings confirm that the CMF method captures essential frustration effects and provides reliable estimates of phase boundaries and thermodynamic quantities. In the following, we employ CMF calculations using 6-site and 18-site clusters to characterize the phase transitions in the frustrated honeycomb Ising model near the boundary separating the ferromagnetic and antiferromagnetic phases.

The remainder of this paper is organized as follows.
Section~\ref{sec:model} introduces the model and the CMF method. Section~\ref{sec:res} presents our results along with their discussion. Finally, Section~\ref{sec:conc} summarizes the main conclusions and offers an outlook for future relevant research directions.

\begin{figure}
\centering
\includegraphics[width=0.97\linewidth]{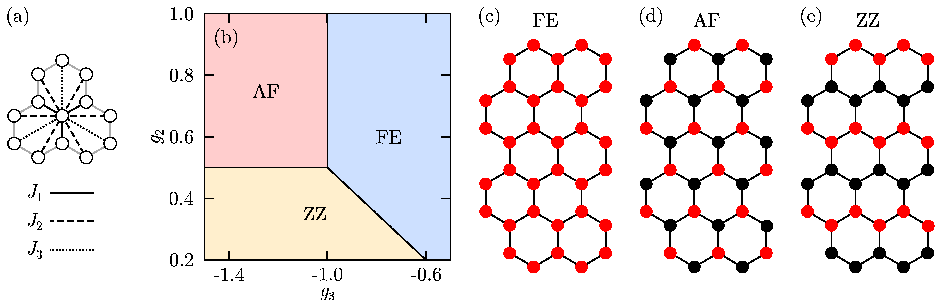}
\caption{(a) Schematic representation of the first-, second-, and third-neighbor interactions in the honeycomb lattice.
(b) Portion of the ground-state phase diagram in the $(g_2, g_3)$ plane, where $g_2 = J_2/J_1$ and $g_3 = J_3/J_1$.
Panels (c)--(e) illustrate the magnetic configurations of the ferromagnetic (FE), N\'eel antiferromagnetic (AF), and zigzag (ZZ) antiferromagnetic phases, respectively. Red and black circles denote spins with opposite orientations.}\label{DiagFund}
\end{figure}

\section{Model and method}
\label{sec:model}

The Ising Hamiltonian in the absence of external fields is given by
\begin{equation}
\mathcal{H} = -\sum_{i,j} J_{ij} \sigma_i \sigma_j, \label{Ham_original}
\end{equation}
where the sum runs over all spin pairs $(i,j)$ coupled by exchange interactions $J_{ij}$, and $\sigma_i = \pm 1$.
Here, we consider a honeycomb lattice with first-, second-, and third-neighbor interactions denoted by $J_1$, $J_2$, and $J_3$, respectively [see Fig.~\ref{DiagFund}(a)].
Depending on the magnitude and sign of these couplings, the ground state of the model can exhibit several distinct forms of long-range order~\cite{PTP_1976}. Previous studies have focused on the phase transitions in the absence of third-neighbor couplings~\cite{SCHMIDT2021168151,Zukovic-2021,Entropy_2024}, a regime where strong frustration effects arise~\cite{SCHMIDT2021168151}. Phase transitions in the case of ferromagnetic $J_1$ and $J_3$ have also been investigated using CMF and Monte Carlo methods~\cite{Pietro2024-MC}, revealing the emergence of three distinct long-range ordered states. For ferromagnetic $J_1$ and antiferromagnetic $J_3$, recent studies focusing on $-1 < g_3 < 0$ have identified zigzag phase transitions~\cite{Pietro2023-PRB}, as illustrated in Fig.~\ref{DiagFund}(b). However, theoretical descriptions of the AF--PM transitions in the model with ferromagnetic $J_1$ and $J_2$ and antiferromagnetic $J_3$ remain scarce. In this work, we address this scenario, concentrating on the phase transitions occurring for $g_2 > 0.5$. As shown in the ground-state phase diagram of Fig.~\ref{DiagFund}(b), this regime is characterized by a competition between ferromagnetic and antiferromagnetic phases, illustrated in panels (c) and (d) of Fig.~\ref{DiagFund}.

The impact of competing interactions on the stability of each magnetic phase can be understood from their respective ground-state energies. For instance, the ground-state energy per spin of the ferromagnetic and antiferromagnetic phases are given by $U_{\textrm{FE}}=-(3J_1+6J_2+3J_3)/2$ and $U_{\textrm{AF}}=(3J_1-6J_2+3J_3)/2$, showing that antiferromagnetic third-neighbor interactions increase the energy of the ferromagnetic phase while lowering that of the antiferromagnetic phase, thereby favoring the latter.
Additionally, the zigzag phase depicted in Fig.~\ref{DiagFund}(e) has a ground-state energy per spin
$U_{\textrm{ZZ}} = -(J_1 - 2J_2 - 3J_3)/2$,
which restricts its stability to $g_2 < 0.5$.
Comparing the ground-state energies of these phases allows the construction of the phase diagram shown in Fig.~\ref{DiagFund}(b).

To investigate the effect of thermal fluctuations on the competition between antiferromagnetic and ferromagnetic phases, we employ the CMF approximation applied to Eq.~\eqref{Ham_original}. Although this mean-field-like method usually overestimates critical temperatures, it also introduces a significant improvement over the single-site mean-field treatment \cite{kerrai2024study} by incorporating short-range correlations \cite{Pietro2023-PRB}.
In this approach, the lattice is divided into $N_{\rm c}$ identical clusters, each containing $n_{\rm s}$ sites, so that the total number of lattice sites is $N = n_{\rm s} N_{\rm c}$.
A mean-field decoupling is then applied between clusters: intracluster interactions are treated exactly, while the intercluster interactions are approximated using the standard mean-field approach,
\begin{equation}
\sigma_i \sigma_j \approx \sigma_i m_j + \sigma_j m_i - m_i m_j,
\end{equation}
where $m_i = \langle \sigma_i \rangle$ denotes the thermal average of the spin at site $i$.

\begin{figure}
\centering
\includegraphics[width=0.75\linewidth]{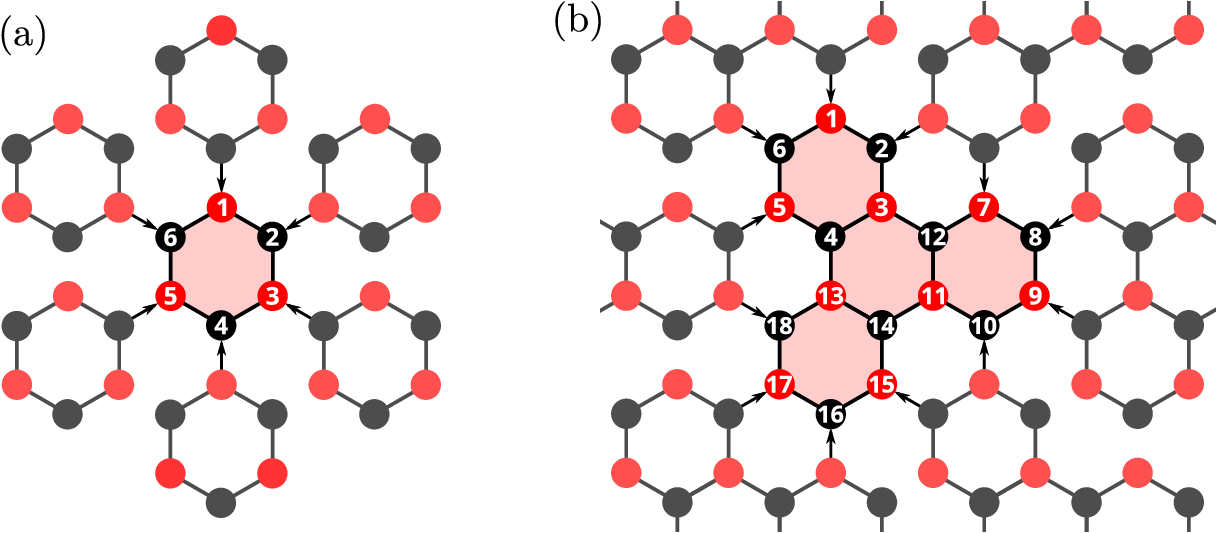}
\caption{Schematic representation of the CMF method applied to the honeycomb lattice using clusters of (a) 6 and (b) 18 sites.
Second- and third-neighbor interactions are omitted for clarity.
Arrows indicate the influence of the mean fields on the central cluster (hatched in red). Red and black filled circles denote spins with opposite directions.} \label{cluster}
\end{figure}

The single-cluster Hamiltonian within the CMF approximation consists of two main contributions: an intracluster term ($H^{\rm c}$), which accounts for the interactions within each cluster, and an intercluster mean-field term ($H_{\textrm{MF}}$), which incorporates the effective fields arising from neighboring clusters. Thus, the CMF Hamiltonian can be written as
\begin{equation}
    H_{\textrm{CMF}}= H^{\rm c} + H_{\textrm{MF}} = -\sum_{(i,j)_c} J_{ij} \sigma_i \sigma_j - \sum_{(i,j)_{c^{'}}} J_{ij} \big( 2\sigma_i - m_i \big) \frac{m_j}{2}, \label{eq:cmf_ham}
\end{equation}
 where the summation index $(i,j)_c$ denotes sites $i$ and $j$ belonging to the same cluster, while $(i,j)_{c^{'}}$ denotes sites $i$ and $j$ located in different clusters.
The term $H_{\textrm{MF}}$ depends on the set of local magnetizations,
\begin{equation}
    m_i= \langle \sigma_i \rangle= \frac{\textrm{Tr} \, \sigma_i e^{-\beta H_{\textrm{CMF}}}}{Z},\label{eq:local_mag}
\end{equation}
where $\beta = 1/(k_{\mathrm{B}} T)$, with $k_{\mathrm{B}}$ being the Boltzmann constant and $T$ the temperature, and
$Z = \mathrm{Tr}\, e^{-\beta H_{\textrm{CMF}}}$ denotes the partition function.
The specific form of the intracluster Hamiltonian and the number of independent local magnetizations depend on the cluster geometry adopted.
In this work, we employ clusters with $n_{\rm s} = 6$ and $n_{\rm s} = 18$ sites, as illustrated in Fig.~\ref{cluster}. These cluster configurations have been used in previous CMF studies of the $J_1$–$J_2$–$J_3$ Ising model on the honeycomb lattice, yielding results in excellent agreement with Monte Carlo simulations~\cite{Pietro2024-MC}. In the following, we present the details of the CMF formulation for these clusters when both ferromagnetic and antiferromagnetic phases are considered.

By adopting 6-site clusters in the CMF approach for the honeycomb lattice (see panel~(a) of Fig.~\ref{cluster}),
the term computed exactly in the Hamiltonian is given by
\begin{equation}
\begin{aligned}
    H^{\rm c}_6 = &-J_1 [\sigma_1(\sigma_2+\sigma_6) +\sigma_3(\sigma_2+\sigma_4) +\sigma_5(\sigma_4+\sigma_6)] -J_2 [\sigma_1 (\sigma_3+\sigma_5) +\sigma_3\sigma_5 +\sigma_2(\sigma_4+\sigma_6) +\sigma_4\sigma_6] \\
    &\quad - J_3[\sigma_1\sigma_4 +\sigma_2\sigma_5 +\sigma_3\sigma_6].
\end{aligned}
\end{equation}
For this cluster, all sites are topologically equivalent, and for the phases under consideration the MF contribution can be expressed in terms of a single local magnetization. In the antiferromagnetic phase, the magnetizations of nearest-neighbor sites differ only by a sign, such that
$m_1=-m_2=m_3=-m_4=m_5=-m_6$,
whereas in the ferromagnetic phase all sites have identical magnetizations,
$m_1=m_2=m_3=m_4=m_5=m_6$.
Accordingly, the mean-field contribution to the Hamiltonian for the 6-site cluster can be written in a unified form as
\begin{equation}
    H_{\textrm{MF}-6}^{\alpha} = \mathcal{J}_\alpha (\mathcal{S}_\alpha -3m_{\alpha}) m_{\alpha},
\end{equation}
where $\alpha =$ AF or FE.
For the antiferromagnetic phase,
$\mathcal{J}_{\mathrm{AF}} = J_1 - 4J_2 + 2J_3$ and $\mathcal{S}_{\mathrm{AF}} = \sum_{i=1}^{6} (-1)^{i-1}\sigma_i$,
while for the ferromagnetic phase,
$\mathcal{J}_{\mathrm{FE}} = -J_1 - 4J_2 - 2J_3$ and $\mathcal{S}_{\mathrm{FE}} = \sum_{i=1}^{6} \sigma_i$. In addition, the order parameters are given by $m_{\alpha} = \langle \mathcal{S}_{\alpha}\rangle/6$.
Thus, the CMF Hamiltonians for the two phases are written as
\begin{align}
    H_{\mathrm{CMF}\text{-}6}^{\mathrm{AF}} &= H^c_6 + H_{\mathrm{MF}\text{-}6}^{\mathrm{AF}}, \\
    H_{\mathrm{CMF}\text{-}6}^{\mathrm{FE}} &= H^c_6 + H_{\mathrm{MF}\text{-}6}^{\mathrm{FE}}.
\end{align}

Since the Hamiltonian within the 6-site CMF approach can be expressed in terms of a single self-consistent parameter, which also serves as the order parameter for both ferromagnetic and antiferromagnetic phases,
the nature of the order–disorder phase transitions can be analyzed through the derivatives of the free energy per cluster,
\begin{equation}
    f_{\alpha}=-k_BT \ln \textrm{Tr} \, e^{- \beta H_{\textrm{CMF}-n_{\rm s}}^{\alpha}}.
\end{equation}
By expanding $f_{\alpha}$ in powers of the order parameter $m_\alpha$, we obtain the following expressions for the second- and fourth-order derivatives:
\begin{equation}
    \eta_2^{\alpha} = \frac{\partial^2f_{\alpha}}{\partial m_\alpha^2} \left. \Bigg| \right. _{m_\alpha=0} = -\mathcal{J}_{\alpha} \Big( 6+ \beta \mathcal{J}_{\alpha} \left\langle \mathcal{S}_{\alpha}^2 \right\rangle_c \Big)
\end{equation}
and
\begin{equation}
\begin{aligned}
    \eta_4^{\alpha} = \frac{\partial^4f_{\alpha}}{\partial m_\alpha^4} \left. \Bigg| \right. _{m_\alpha=0} = \beta^3 \mathcal{J}_{\alpha}^4 \Big( 3\left\langle \mathcal{S}_{\alpha}^2 \right\rangle_{\rm c}^2- \left\langle \mathcal{S}_{\alpha}^4 \right\rangle_{\rm c} \Big),
\end{aligned}
\end{equation}
where $\langle \cdots \rangle_{\rm c}$ denotes the Boltzmann average taken with respect to $H_6^{\rm c}$.
Second-order phase transitions are identified by the condition $\eta_2^{\alpha} = 0$ with $\eta_4^{\alpha} > 0$,
whereas a tricritical point occurs when $\eta_2^{\alpha} = \eta_4^{\alpha} = 0$.

In addition to the 6-site formulation, we also study the model described by Eq.~\eqref{Ham_original} using the CMF approximation within an 18-site cluster (see panel~(b) of Fig.~\ref{cluster}).
In this case, the mean-field contribution $H_{\mathrm{MF}}$ involves three independent self-consistent parameters,
as the larger cluster contains topologically non-equivalent sites.
Due to the extensive form of the resulting equations for the $n_{\rm s} = 18$ case,
the full details of this approximation are presented in Appendix~\ref{hamiltonianos}. For both cluster sizes considered, the order parameters characterizing the antiferromagnetic and ferromagnetic phases are defined as
\begin{equation}
    m_{\mathrm{AF}} = \frac{1}{n_{\rm s}} \sum_{i=1}^{n_{\rm s}} (-1)^{i-1} m_i,
    \qquad
    m_{\mathrm{FE}} = \frac{1}{n_{\rm s}} \sum_{i=1}^{n_{\rm s}} m_i,
\end{equation}
respectively.
We also calculate the entropy per spin,
\begin{equation}
    S = \frac{U - f}{n_{\rm s} T},
\end{equation}
where $U$ is the internal energy per cluster, evaluated as the thermal average of the CMF Hamiltonian~\cite{Pietro2023-PRB},
and $f$ is the free energy per cluster.
The combined analysis of the entropy and order parameters, together with the free energy behavior, allows us to determine the nature of the phase transitions within both levels of the CMF theory.

\section{Results}
\label{sec:res}

In this section, we present the results obtained using the CMF approximation with clusters of 6 and 18 sites. We analyze the critical behavior and phase transitions of the model within the range of exchange interactions where the ferromagnetic and antiferromagnetic phases compete, i.e., for $g_2 > 0.5$.
Within the 6-site cluster approach, the nature of the order–disorder phase transitions is determined through the analysis of the free-energy derivatives and landscapes, $\Delta f = f_{\textrm{AF}} - f_{\textrm{PM}}$.
For the larger cluster ($n_{\rm s} = 18$), the phase transitions are characterized by examining the behavior of thermodynamic quantities such as the free energy, order parameters, and entropy.

\begin{figure}
\centering
\includegraphics[width=0.95\linewidth]{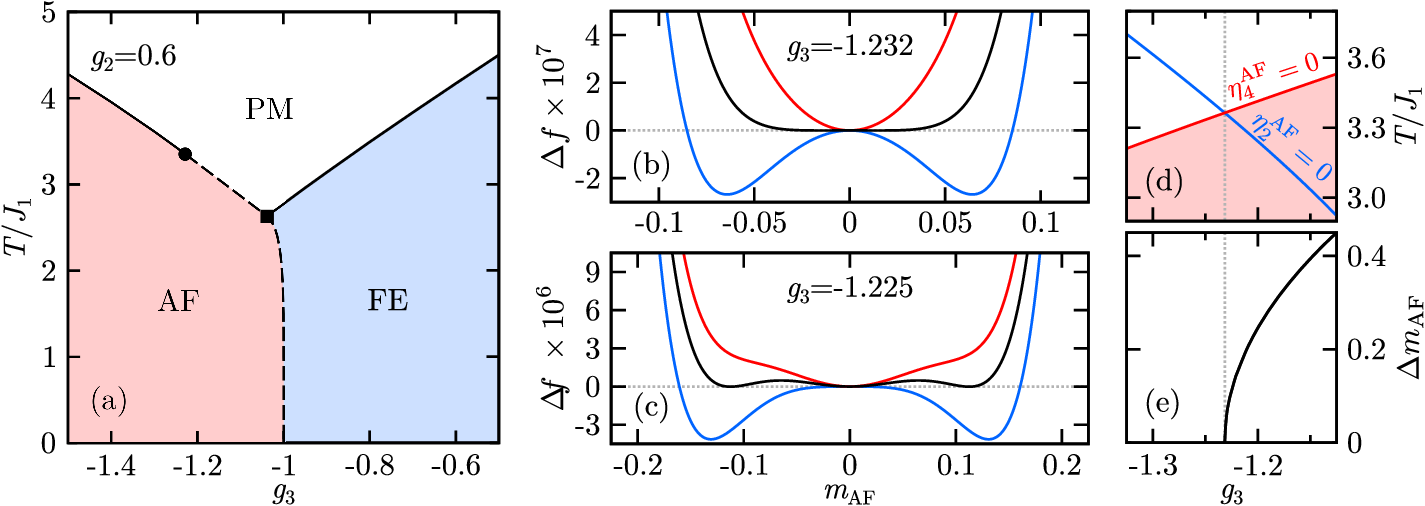}
\caption{Phase transitions for $g_2 = 0.6$ within the 6-site CMF approximation.
(a) Phase diagram of temperature versus third-neighbor interaction $g_3$, where solid and dashed lines represent second- and first-order phase transitions, respectively. The circle marks a tricritical point, and the square denotes a critical endpoint. Panels (b) and (c) show the free-energy difference $\Delta f = f_{\textrm{AF}} - f_{\textrm{PM}}$ for values of $g_3$ near the tricritical point.
The red, black, and blue curves in panel (b) correspond to $T/J_1 = 3.365500$, $3.365443$, and $3.365400$, respectively, while panel (c) shows results for $T/J_1 = 3.339000$ (red), $3.338915$ (black), and $3.338000$ (blue). Panel (d) presents the loci of $\eta^{\textrm{AF}}_2 = 0$ (blue) and $\eta^{\textrm{AF}}_4 = 0$  (red) in the $(g_3, T)$ plane; the light-red and white regions correspond to $\eta^{\textrm{AF}}_4 < 0$ and $\eta^{\textrm{AF}}_4 > 0$, respectively.
(e) Discontinuity of the antiferromagnetic order parameter as a function of $g_3$ at the AF–PM transition. The dotted lines are guides to the eye.}\label{Diag_J2=0.6}
\end{figure}

In Fig.~\ref{Diag_J2=0.6}, we present the phase diagram in the temperature {\it versus} $g_3$ plane for $g_2 = 0.6$, together with the results obtained from different free-energy analyses used to evaluate the AF–PM phase transitions within the 6-site CMF approach. As shown in panel (a), only second-order phase transitions are found between the ferromagnetic and paramagnetic phases.
In contrast, the AF–PM phase boundary exhibits first-order phase transitions for $-1.23 < g_3 < -1.04$. For stronger antiferromagnetic third-neighbor interactions ($g_3 < -1.23$), the transition from the antiferromagnetic to the disordered phase becomes second-order upon heating. The change in the nature of the AF–PM transition can be identified from the behavior of the free-energy difference $\Delta f$ as a function of the antiferromagnetic order parameter $m_{\textrm{AF}}$.
Figure~\ref{Diag_J2=0.6}(b) shows the CMF results for $g_3 = -1.232$, where two symmetric minima are observed for temperatures below the Néel temperature $T_N$, a flat minimum appears near the transition temperature, and a single minimum is found above $T_N$—a typical signature of a second-order phase transition. For slightly weaker antiferromagnetic third-neighbor coupling, a qualitative change occurs in the free-energy landscape. As shown in panel (c), for $g_3 = -1.225$, a three-minima structure emerges, indicating the coexistence of ordered and disordered phases at the transition temperature, as expected for a first-order transition.
This change in the transition character signals the presence of a tricritical point \cite{PhysRevLett.24.715} at the AF–PM phase boundary, marked by a filled circle in the phase diagram.

A precise determination of the tricritical point within the 6-site CMF approach can be achieved by analyzing the behavior of the derivatives of the free energy with respect to the order parameter. Panel (d) of Fig.~\ref{Diag_J2=0.6} displays the curves corresponding to $\eta_2^{\textrm{AF}}=0$ and $\eta_4^{\textrm{AF}}=0$ in the $T/J_1$–$g_3$ plane. The locus of points satisfying $\eta_2^{\textrm{AF}}=0$ marks the AF–PM transition temperature across the $g_3$ range when $\eta_4^{\textrm{AF}}>0$, coinciding with the second-order phase transition line shown in panel (a). The intersection between the $\eta_2^{\textrm{AF}}=0$ and $\eta_4^{\textrm{AF}}=0$ curves defines the coordinates of the tricritical point ($g_3 \approx -1.228$), indicated by a filled circle in panel (a).
To further substantiate the analysis based on free-energy derivatives, panel (e) presents the discontinuity of the antiferromagnetic order parameter, $\Delta m_{\textrm{AF}}$, as a function of the third-neighbor coupling $g_3$. The quantity $\Delta m_{\textrm{AF}}$ vanishes precisely at the same $g_3$ value where the $\eta_2^{\textrm{AF}}=0$ and $\eta_4^{\textrm{AF}}=0$ curves intersect, confirming that the first-order AF–PM transition terminates at the tricritical point as $g_3$ decreases. Additionally, the transitions between the antiferromagnetic and ferromagnetic phases are found to be of first order, as determined by comparing the free energies of these phases. The second-order FE–PM phase boundary ends at a critical endpoint~\cite{Stishov2020}, represented by a filled square in the phase diagram, where it meets the line of first-order transitions delimiting the antiferromagnetic phase.

It is worth noting that tricriticality is the subject of several recent studies~\cite{Moueddene_2024, PhysRevE.110.064144, PhysRevResearch.7.013214, PhysRevResearch.7.033240, ROCHANETO2023129145, saadi2024critical} and has been reported in other frustrated Ising systems with competing interactions, including monolayer and bilayer square and honeycomb lattices~\citep{Pietro2023-PRB,Pietro2024-MC,Roos2024,Rossato2023,Dominguez2021-CV,Hu2021-TM,PhysRevE.109.014144,Chatelain2025-TN, ROOS2026131093}. However, previous studies of the $J_1$–$J_2$ Ising model on the honeycomb lattice have found only second-order phase transitions for $ -1/4 < g_2 < 0 $~\cite{SCHMIDT2021168151,Entropy_2024,Zukovic-2021}. Therefore, the inclusion of an antiferromagnetic third-neighbor coupling plays a crucial role not only in stabilizing the antiferromagnetic phase but also in driving the emergence of tricriticality.
These findings are particularly relevant in light of the growing interest in van der Waals magnetic materials, which can be synthesized as genuine two-dimensional spin systems~\citep{Review2022-2DMaterials,Review2022-VanDerWaals,Park2016-VanDerWaals}, with several compounds exhibiting strong Ising anisotropy~\citep{FePS3-2016,VBr3-2023}. In the following, we analyze how varying the strength of $g_2$ influences the extent of the first-order AF–PM transition regime and assess the robustness of our results by comparing the 6-site and 18-site CMF approximations.

\begin{figure}
\centering
\includegraphics[width=0.95\linewidth]{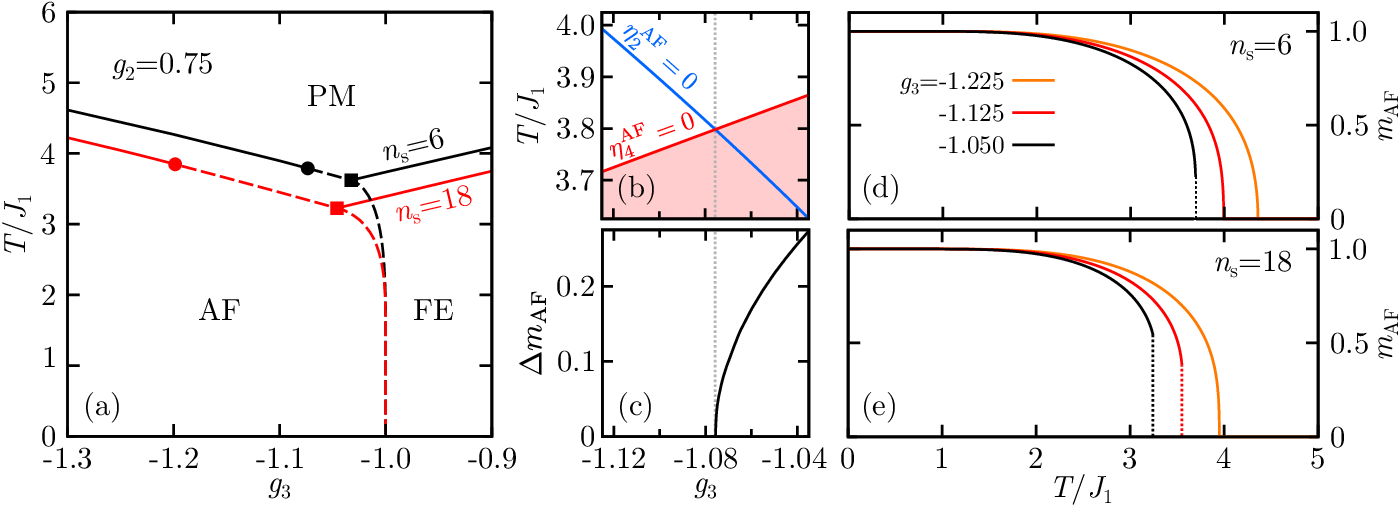}
\caption{Phase transitions for $g_2 = 0.75$.
(a) Temperature–coupling phase diagrams obtained within the CMF approach using clusters of 6 (dark) and 18 (red) sites. (b) Curves satisfying $\eta_2^{\textrm{AF}} = 0$ and $\eta_4^{\textrm{AF}} = 0$ in the temperature–$g_3$ plane for $n_s=6$. The regions with $\eta_4^{\textrm{AF}} < 0$ and $\eta_4^{\textrm{AF}} > 0$ are indicated in light red and white, respectively.
(c) Discontinuity of the antiferromagnetic order parameter for $n_{\rm s} = 6$ as a function of $g_3$. Panels (d) and (e) show the temperature dependence of the antiferromagnetic order parameter obtained from the CMF approximation with $n_{\rm s} = 6$ and $n_{\rm s} = 18$, respectively.}
\label{Diag_J2=0.75}
\end{figure}

Figure~\ref{Diag_J2=0.75} presents the results for $g_2 = 0.75$, including the phase diagram in the temperature--$g_3$ plane for $n_{\rm s} = 6$ and $n_{\rm s} = 18$, shown in panel (a). Although first-order phase transitions are still observed along the AF--PM phase boundary, the range of $g_3$ values for which they occur for $n_{\rm s}=6$ is considerably reduced compared with the results for $g_2 = 0.60$ displayed in Fig.~\ref{Diag_J2=0.6}(a). This behavior indicates that increasing the strength of the ferromagnetic second-neighbor interactions promotes the emergence of second-order AF--PM phase transitions. Panel (b) displays the curves satisfying $\eta_2^{\textrm{AF}} = 0$ and $\eta_4^{\textrm{AF}} = 0$, whose intersection at $g_3 \approx -1.074$ marks the location of the tricritical point in the phase diagram. This result is consistent with the behavior of the antiferromagnetic order parameter discontinuity, which also vanishes at the same value of $g_3$, as shown in panel~(c). By employing the 18-site cluster in the CMF approximation, the phase diagram for $g_2 = 0.75$ [shown in red in Fig.~\ref{Diag_J2=0.75}(a)] also displays a region of first-order phase transitions between the antiferromagnetic and paramagnetic phases, bounded by a tricritical point and a critical endpoint.
However, the range of $g_3$ values over which these first-order transitions occur is broader than that obtained with the 6-site cluster approach, whose results are shown in black in the same panel.

To highlight the differences between the two levels of approximation, panels (d) and (e) of Fig.~\ref{Diag_J2=0.75} display the antiferromagnetic order parameter as a function of temperature for $n_{\rm s} = 6$ and $n_{\rm s} = 18$, respectively. For $g_3 = -1.050$, both cluster sizes indicate a first-order phase transition, evidenced by the discontinuity in $m_{\textrm{AF}}$. In contrast, for $g_3 = -1.125$, the 6-site cluster approach predicts a second-order phase transition, whereas the 18-site cluster yields a first-order one. These findings suggest that increasing the cluster size extends the range of third-neighbor interactions over which first-order AF–PM transitions occur. Nevertheless, both approximations show qualitative agreement for $g_3 = -1.225$, where second-order AF–PM transitions are observed. It is also worth noting that the larger cluster systematically produces lower transition temperatures, a direct consequence of the improved treatment of correlations achieved by including more interactions within the exact enumeration.

\begin{figure}
\centering
\includegraphics[width=1.0\linewidth]{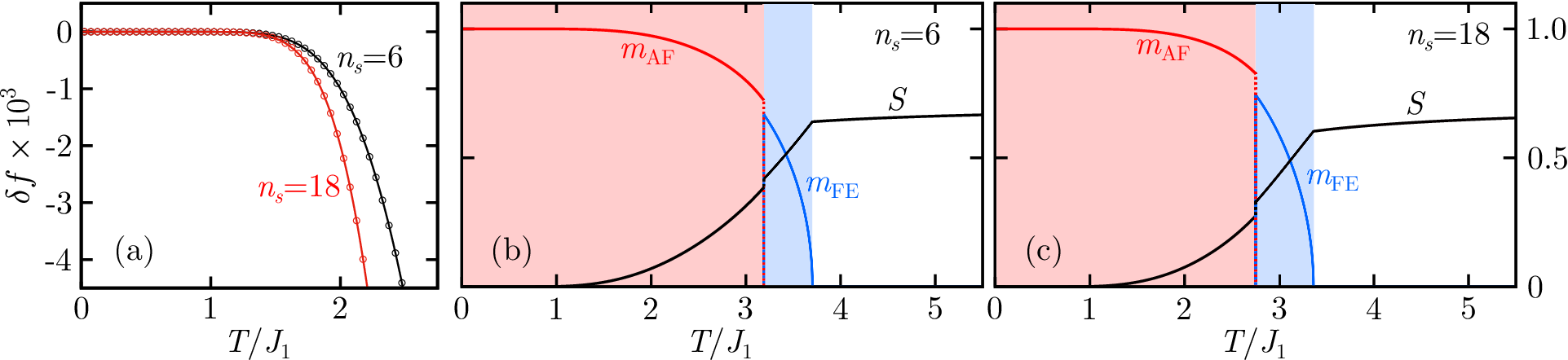}
\caption{(a) Difference between the free energies per spin of the ferromagnetic and antiferromagnetic phases, $\delta f = (f_{\textrm{FE}} - f_{\textrm{AF}})/n_{\rm s}$, as a function of temperature for $g_2 = 0.75$ and $g_3 = -1.0$, obtained using the 6-site ($n_{\rm s} = 6$, dark) and 18-site ($n_{\rm s} = 18$, red) CMF approximations. Panels (b) and (c) show the temperature dependence of the entropy per spin and the order parameters of the antiferromagnetic and ferromagnetic phases for $g_2 = 0.75$ and $g_3 = -1.01$ within the 6-site and 18-site CMF approaches, respectively.}
\label{reentrancia}
\end{figure}

The competition between interactions leads to a direct AF--FE transition at $g_3 = -1$ in the absence of thermal fluctuations. When temperature is included, interesting phenomena emerge in the regime of strong competition between these long-range orders. Figure~\ref{reentrancia}(a) shows the difference between the free energies of the antiferromagnetic and ferromagnetic phases as a function of temperature for $g_2 = 0.75$ and $g_3 = -1.0$, computed using the $n_{\rm s} = 6$ and $n_{\rm s} = 18$ CMF approaches. At low temperatures, the free energies of both phases are equal, indicating coexistence. As the temperature increases, the free energy of the ferromagnetic phase decreases more rapidly than that of the antiferromagnetic phase, causing $\delta f = (f_{\textrm{FE}} - f_{\textrm{AF}})/n_{\rm s}$ to become negative. This behavior demonstrates that thermal fluctuations favor the ferromagnetic phase over the antiferromagnetic phase, revealing an order-by-disorder state selection. This phenomenon can be attributed to the entropic selection of a particular phase in a degenerate scenario, where the phase with the highest density of closer excited states is entropically favored \cite{Albarracin2018, bergman2007order, OBD_1980}. Notably,  the entropic selection of the FE phase driven by thermal fluctuations is consistently observed in both cluster sizes, $n_{\rm s} = 6$ and $n_{\rm s} = 18$, within the present model.

We also observe transitions between the ordered phases as the temperature increases for fixed values of $g_3$ slightly below $-1.0$. This behavior is illustrated in panels (b) and (c) of Fig. \ref{reentrancia}, which display the order parameters and entropy per spin as functions of temperature for $n_{\rm s} = 6$ and $n_{\rm s} = 18$, respectively, with $g_2 = 0.75$ and $g_3 = -1.01$. The first-order AF--FE transition is indicated by the discontinuous change in the entropy, occurring at the boundary between the red (AF) and blue (FE) regions. This transition is also evident in the order parameters: $m_{\textrm{AF}}$ drops discontinuously to zero at the same point where $m_{\textrm{FE}}$ becomes nonzero. Upon further increasing the temperature, the system undergoes a second-order FE--PM transition. These two successive transitions near $g_3 = -1$ are observed over a wide range of $g_2$. Notably, both the order-by-disorder state selection and the occurrence of two successive transitions are consistently captured in both CMF approaches considered here, and similar phenomena have been reported in other spin systems with competing interactions~\cite{Pietro2023-PRB, Rossato2023,Albarracin2018}.

\begin{figure}
\centering
\includegraphics[width=0.96\linewidth]{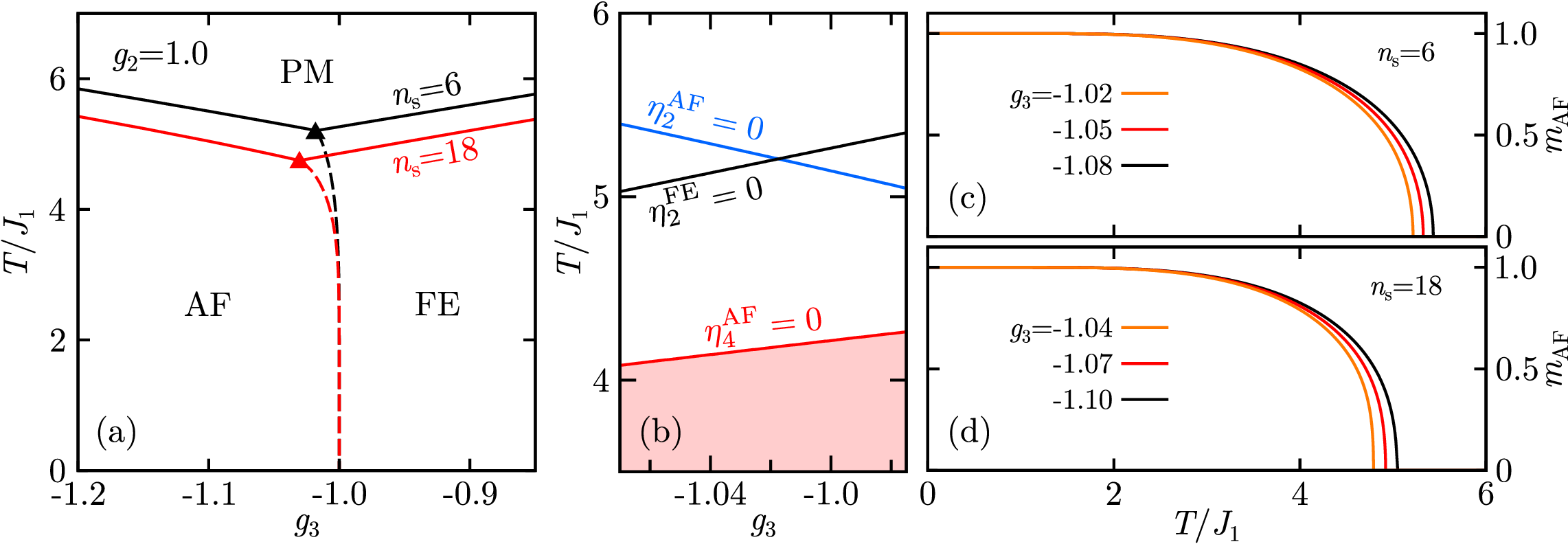}
\caption{Phase transitions for $g_2 = 1.0$. (a) Temperature--$g_3$ phase diagram obtained within the 6-site and 18-site CMF approaches, where the triangles indicate bicritical points. (b) Curves satisfying $\eta_2^{\textrm{FE}} = 0$, $\eta_2^{\textrm{AF}} = 0$, and $\eta_4^{\textrm{AF}} = 0$, where the region with $n_4^{AF}<0$ is indicated in light red color. Panels (c) and (d) show the temperature dependence of the antiferomagnetic order parameter near the bicritical point for $n_{\rm s} = 6$ and $n_{\rm s} = 18$, respectively.}
\label{Diag_J2=1.0}
\end{figure}

As evident from a comparison between Figs.~\ref{Diag_J2=0.6} and \ref{Diag_J2=0.75}, increasing the second-neighbor interactions reduces the range of first-order AF--PM transitions, raising the question of whether sufficiently large $g_2$ can suppress these first-order transitions entirely. Motivated by this, we examine phase diagrams for stronger second-neighbor interactions and find that, when $g_2$ is large enough, the system exhibits only second-order phase transitions in the order-disorder  boundary. This scenario is illustrated in Fig.~\ref{Diag_J2=1.0}, which shows the phase diagrams for $g_2 = 1.0$ obtained using the 6- and 18-site CMF approaches (panel (a)). In this regime, where second-neighbor interactions are comparable to first-neighbor interactions, tricriticality is absent.
However, the phase diagram exhibits a bicritical point \cite{PhysRevLett.32.1350}, indicated by a triangle, where the two second-order transition lines meet the first-order FE--AF transition line~\cite{Stishov2020}.  This special point differs from the tricritical point, as no order-disorder first-order phase transition is present and there are two ordered phases associated with the bicritical point. Within the $n_{\rm s} = 6$ approach, the bicritical point can be identified as the intersection of the curves $\eta_2^{\textrm{AF}} = 0$ and $\eta_2^{\textrm{FE}} = 0$ (panel (b)), which mark the order-disorder boundaries in their respective regimes. The antiferromagnetic order parameter further supports this interpretation. Panels (c) and (d) show $m_{\textrm{AF}}$ as a function of temperature for the 6- and 18-site approaches, respectively, for $g_3$ values near the bicritical point, revealing continuous behavior. These results demonstrate that sufficiently strong second-neighbor interactions can suppress first-order AF--PM transitions. This change in the nature of phase transitions can be attributed to the reduction of frustration effects driven by the increase in the ferromagnetic second-neighbor interactions. The first-order AF--PM transitions can be seen as a consequence of the frustration driven by the competition between first- and third-neighbor interactions, which can be mitigated by the increase in $g_2$ that also increases the order-disorder transition temperature.

It is worth noting that all FE--PM phase transitions observed in this work are second-order. Our CMF results, at both levels of approximation and across a wide range of parameters—from the highly frustrated region near the AF--FE--ZZ triple point shown in Fig.~\ref{DiagFund} to values of $g_2$ comparable to $|g_3|$—consistently indicate second-order FE--PM transitions. According to the ground-state phase diagram of the model~\cite{WildesPRB-2020}, the ferromagnetic phase can share boundaries with different antiferromagnetic phases. Recent CMF studies also show that, regardless of the values of second- and third-neighbor interactions, second-order FE--PM transitions persist whenever the first-neighbor interactions are ferromagnetic~\cite{Pietro2023-PRB,Pietro2024-MC}. Therefore, based on the present study, we conclude that the continuous character of the FE--PM transitions is a robust feature of the $J_1$-$J_2$-$J_3$ Ising model on the honeycomb lattice.

\begin{figure}
\centering
\includegraphics[width=0.97\linewidth]{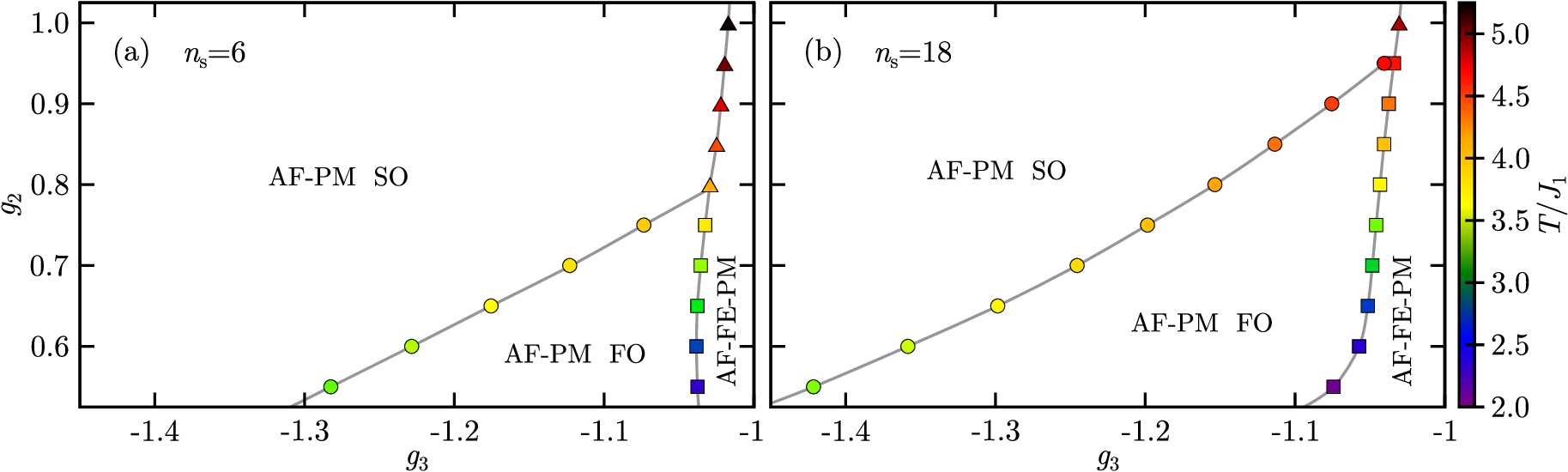}
\caption{Coordinates of the critical endpoints (squares), bicritical points (triangles), and tricritical points (circles), with the color scale indicating the corresponding temperatures. Panels (a) and (b) show results obtained using the 6-site and 18-site CMF approaches, respectively. The gray curves are guides to the eye, highlighting regions of first- (FO) and second-order (SO) phase transitions, as well as the successive AF--FE--PM transitions induced by thermal fluctuations.}
\label{PontosTri}
\end{figure}

In Fig.~\ref{PontosTri}, we present an overview of the phase transitions obtained within the CMF approximation for $n_{\rm s} = 6$ and $18$, shown in panels (a) and (b), respectively. The coupling coordinates of tricritical, bicritical, and critical endpoints are indicated by filled circles, triangles, and squares, with colors representing the corresponding temperatures.
The region between tricritical and critical endpoints corresponds to AF--PM first-order transitions. This region is larger for $n_{\rm s} = 18$, suggesting that the onset of first-order AF--PM transitions is not an artifact of the mean-field-like approximation. Increasing $g_2$ suppresses the first-order AF--PM transitions, highlighting the role of ferromagnetic second-neighbor interactions in eliminating tricriticality. We also identify a region in the $g_2$--$g_3$ plane where two successive phase transitions occur as temperature increases. This region lies between $g_3 = -1.0$ and the critical endpoints (for smaller $g_2$) or the bicritical points (for larger $g_2$), corresponding to transitions analogous to those shown in Fig.~\ref{reentrancia}. Based on the trend of the bicritical and critical endpoints, this phenomenon is expected to persist for even larger $g_2$.
It is worth emphasizing that the order-by-disorder phenomenon accompanies the occurrence of successive phase transitions due to the shift of the coupling coordinates of the bicritical points and critical endpoints relative to the AF-FE ground-state phase boundary.
As a general trend, the temperature coordinates of these special points increase with $g_2$, as indicated by the color scale on the right.

\section{Summary and outlook}
\label{sec:conc}

In this work, we investigated the phase transitions hosted by the Ising model on the honeycomb lattice with first-, second-, and third-neighbor interactions. We considered ferromagnetic $J_1$ and $J_2$, and antiferromagnetic $J_3$, focusing our analysis on the vicinity of the boundary between the antiferromagnetic and ferromagnetic phases. Using the cluster mean-field approximation with clusters of 6 and 18 sites, we examined the thermal phase transitions exhibited by the model. The behavior of thermodynamic quantities, such as the order parameter and entropy, together with the free-energy landscape and its derivatives (for $n_{\rm s} = 6$), were used to characterize the nature of the transitions.

Our main findings indicate that both first- and second-order phase transitions, as well as tricritical behavior, occur along the AF--PM phase boundary. As the ferromagnetic second-neighbor interaction increases, the range of third-neighbor interactions that support first-order AF--PM phase transitions becomes progressively narrower, until only second-order phase transitions remain. Notably, the region of first-order AF--PM transitions is considerably larger for the $n_{\rm s} = 18$ cluster, indicating that the occurrence of this type of transition is robust upon increasing the cluster size. In agreement with previous studies of the model~\cite{SCHMIDT2021168151,Pietro2023-PRB,Pietro2024-MC}, we found exclusively second-order FE--PM phase transitions. Furthermore, near the AF--FE phase boundary, two successive phase transitions may occur as the temperature increases: a first-order AF--FE transition followed by a second-order FE--PM transition. This behavior arises in regimes of strong competition between the ordered phases. Additionally, for $g_3 = -1.0$, we observed an order-by-disorder mechanism, in which the ferromagnetic phase is stabilized by thermal fluctuations.

We believe that the competition between antiferromagnetic and ferromagnetic phases in the frustrated $J_1$–$J_2$–$J_3$ Ising model on the honeycomb lattice deserves further investigation through complementary analytical and numerical techniques. Approaches such as Monte Carlo simulations~\cite{Zukovic-2021,Moueddene_2024, PhysRevResearch.7.013214,PhysRevE.110.064144, ZAIM2024113536,PSS_2025}, cluster variational methods~\cite{VariacionalPRE-2015}, effective-field theory~\cite{EfetivoPRB-2021}, tensor-network approaches~\cite{TensorPRR-2021}, and transfer-matrix formalisms~\cite{MatrizPRB-2021,PhysRevResearch.7.033240} can provide deeper insight into the phase diagrams and critical behavior of this frustrated system. An especially intriguing direction concerns the influence of an external magnetic field on the antiferromagnetic phase transitions, as the possible emergence of field-induced tricriticality remains an open and compelling question. In this context, recent advances in van der Waals materials such as VBr$_3$~\cite{VBr3-2023}, FePS$_3$~\cite{WildesPRB-2020,Pietro2023-PRB}, and FePSe$_3$~\cite{PhysRevB.109.134410} suggest that competing exchange interactions are indeed present in real honeycomb-lattice magnets, thereby motivating further exploration of both quantum and thermal effects within the present model.

\section*{Acknowledgments}
PFD,  FMZ, and MS acknowledge the support from CNPq/Brazil. FMZ also acknowledges support from FUNDECT-MS (Brazil).
The work of NGF was supported by the Engineering and Physical
Sciences Research Council (Grant No. EP/X026116/1).
\appendix

\section{18-site cluster mean-field theory}
\label{hamiltonianos}

In this Appendix, we present the details of the cluster mean-field approach applied to a system divided into clusters containing 18 sites each [see Fig.~\ref{cluster}(b)]. After performing the mean-field decoupling, the cluster mean-field Hamiltonian takes the same general form as Eq.~\eqref{eq:cmf_ham},
with the intracluster contribution given by
\begin{equation}
\begin{aligned}
    H_{18}^{\rm c} &= -J_1[\sigma_1 (\sigma_2 +\sigma_6 )+\sigma_3 (\sigma_2 +\sigma_4 +\sigma_{12} )+\sigma_5 (\sigma_4 +\sigma_6 )+\sigma_7 (\sigma_8 +\sigma_{12} ) +\sigma_9 (\sigma_8 +\sigma_{10} ) \\
    &\quad\quad +\sigma_{11} (\sigma_{10} +\sigma_{12} +\sigma_{14} )+\sigma_{13} (\sigma_4 +\sigma_{14} +\sigma_{18} ) +\sigma_{15} (\sigma_{14} +\sigma_{16} )+\sigma_{17} (\sigma_{16} +\sigma_{18} )] \\
    &\quad -J_2[ \sigma_2 (\sigma_6 +\sigma_{12}) +\sigma_5 (\sigma_1 +\sigma_{13}) +\sigma_7 (\sigma_3 +\sigma_9) +\sigma_3 (\sigma_1 +\sigma_5 +\sigma_{13}) +\sigma_4 (\sigma_2 +\sigma_6 +\sigma_{12}) \\
    &\quad\quad +\sigma_{10} (\sigma_8 +\sigma_{14}) +\sigma_{15} (\sigma_{11} +\sigma_{17})+\sigma_{18} (\sigma_4 +\sigma_{16}) +\sigma_{12} (\sigma_8 +\sigma_{10} +\sigma_{14})+\sigma_{11} (\sigma_3 +\sigma_7 +\sigma_9) \\
    &\quad\quad +\sigma_{13} (\sigma_{11} +\sigma_{15} +\sigma_{17})+\sigma_{14} (\sigma_4 +\sigma_{16} +\sigma_{18}) ]
    -J_3[\sigma_2 (\sigma_5 +\sigma_7)+\sigma_3 (\sigma_6 +\sigma_{14}) \\
    &\quad\quad +\sigma_{10} (\sigma_7 +\sigma_{15})+\sigma_{11} (\sigma_4 +\sigma_8) +\sigma_{13} (\sigma_{12} +\sigma_{16}) +\sigma_{18}(\sigma_5+\sigma_{15}) +\sigma_1 \sigma_4 +\sigma_9 \sigma_{12} +\sigma_{14} \sigma_{17} ].
\end{aligned}
\end{equation}
The intercluster contribution to the cluster mean-field Hamiltonian can be expressed as a function of three independent local magnetizations.
For each magnetic phase considered, specific relations among the local magnetizations of topologically equivalent sites can be established,
which allows us to write the intercluster Hamiltonian in the general form
\begin{equation}
\begin{aligned}
    H^{\textrm{FE/AF}}_{\textrm{MF}-18} =& [3m_1- \kappa_1 \mp \kappa_6][J_1m_2 +J_2(2m_1\pm m_2+m_3) +J_3(m_2\pm m_3)] \\
    &+[3m_2-\kappa_2 \mp \kappa_5][(J_1+J_3)m_1 +J_2(2m_2\pm m_1)] +[3m_3-\kappa_3\mp\kappa_4](J_2\pm J_3)m_1,
\end{aligned}
\end{equation}
where the local magnetizations are given by Eq.~(\ref{eq:local_mag}), and $\kappa_1=\sigma_1+\sigma_9+\sigma_{17}$, $\kappa_2=\sigma_2+\sigma_{10}+\sigma_{18}$, $\kappa_3=\sigma_3+\sigma_{11}+\sigma_{13}$, $\kappa_4=\sigma_4+\sigma_{12}+\sigma_{14}$, $\kappa_5=\sigma_5+\sigma_7+\sigma_{15}$, and $\kappa_6=\sigma_6+\sigma_8+\sigma_{16}$.
 
\bibliography{cas-refs}

\begin{thebibliography}{51}%
\makeatletter
\providecommand \@ifxundefined [1]{%
 \@ifx{#1\undefined}
}%
\providecommand \@ifnum [1]{%
 \ifnum #1\expandafter \@firstoftwo
 \else \expandafter \@secondoftwo
 \fi
}%
\providecommand \@ifx [1]{%
 \ifx #1\expandafter \@firstoftwo
 \else \expandafter \@secondoftwo
 \fi
}%
\providecommand \natexlab [1]{#1}%
\providecommand \enquote  [1]{``#1''}%
\providecommand \bibnamefont  [1]{#1}%
\providecommand \bibfnamefont [1]{#1}%
\providecommand \citenamefont [1]{#1}%
\providecommand \href@noop [0]{\@secondoftwo}%
\providecommand \href [0]{\begingroup \@sanitize@url \@href}%
\providecommand \@href[1]{\@@startlink{#1}\@@href}%
\providecommand \@@href[1]{\endgroup#1\@@endlink}%
\providecommand \@sanitize@url [0]{\catcode `\\12\catcode `\$12\catcode
  `\&12\catcode `\#12\catcode `\^12\catcode `\_12\catcode `\%12\relax}%
\providecommand \@@startlink[1]{}%
\providecommand \@@endlink[0]{}%
\providecommand \url  [0]{\begingroup\@sanitize@url \@url }%
\providecommand \@url [1]{\endgroup\@href {#1}{\urlprefix }}%
\providecommand \urlprefix  [0]{URL }%
\providecommand \Eprint [0]{\href }%
\providecommand \doibase [0]{http://dx.doi.org/}%
\providecommand \selectlanguage [0]{\@gobble}%
\providecommand \bibinfo  [0]{\@secondoftwo}%
\providecommand \bibfield  [0]{\@secondoftwo}%
\providecommand \translation [1]{[#1]}%
\providecommand \BibitemOpen [0]{}%
\providecommand \bibitemStop [0]{}%
\providecommand \bibitemNoStop [0]{.\EOS\space}%
\providecommand \EOS [0]{\spacefactor3000\relax}%
\providecommand \BibitemShut  [1]{\csname bibitem#1\endcsname}%
\let\auto@bib@innerbib\@empty
\bibitem [{\citenamefont {Zhang}\ \emph {et~al.}(2024)\citenamefont {Zhang},
  \citenamefont {Lu}, \citenamefont {Tabrizian}, \citenamefont {Feng},\ and\
  \citenamefont {Wu}}]{zhang20242d}%
  \BibitemOpen
  \bibfield  {author} {\bibinfo {author} {\bibfnamefont {B.}~\bibnamefont
  {Zhang}}, \bibinfo {author} {\bibfnamefont {P.}~\bibnamefont {Lu}}, \bibinfo
  {author} {\bibfnamefont {R.}~\bibnamefont {Tabrizian}}, \bibinfo {author}
  {\bibfnamefont {P.~X.-L.}\ \bibnamefont {Feng}}, \ and\ \bibinfo {author}
  {\bibfnamefont {Y.}~\bibnamefont {Wu}},\ }\href@noop {} {\bibfield  {journal}
  {\bibinfo  {journal} {npj Spintronics}\ }\textbf {\bibinfo {volume} {2}},\
  \bibinfo {pages} {6} (\bibinfo {year} {2024})}\BibitemShut {NoStop}%
\bibitem [{\citenamefont {Elahi}\ \emph {et~al.}(2022)\citenamefont {Elahi},
  \citenamefont {Dastgeer}, \citenamefont {Nazir}, \citenamefont {Nisar},
  \citenamefont {Bashir}, \citenamefont {Qureshi}, \citenamefont {Kim},
  \citenamefont {Aziz}, \citenamefont {Aslam}, \citenamefont {Hussain},
  \citenamefont {Assiri},\ and\ \citenamefont
  {Imran}}]{Review2022-2DMaterials}%
  \BibitemOpen
  \bibfield  {author} {\bibinfo {author} {\bibfnamefont {E.}~\bibnamefont
  {Elahi}}, \bibinfo {author} {\bibfnamefont {G.}~\bibnamefont {Dastgeer}},
  \bibinfo {author} {\bibfnamefont {G.}~\bibnamefont {Nazir}}, \bibinfo
  {author} {\bibfnamefont {S.}~\bibnamefont {Nisar}}, \bibinfo {author}
  {\bibfnamefont {M.}~\bibnamefont {Bashir}}, \bibinfo {author} {\bibfnamefont
  {H.~A.}\ \bibnamefont {Qureshi}}, \bibinfo {author} {\bibfnamefont
  {D.}~\bibnamefont {Kim}}, \bibinfo {author} {\bibfnamefont {J.}~\bibnamefont
  {Aziz}}, \bibinfo {author} {\bibfnamefont {M.}~\bibnamefont {Aslam}},
  \bibinfo {author} {\bibfnamefont {K.}~\bibnamefont {Hussain}}, \bibinfo
  {author} {\bibfnamefont {M.~A.}\ \bibnamefont {Assiri}}, \ and\ \bibinfo
  {author} {\bibfnamefont {M.}~\bibnamefont {Imran}},\ }\href {\doibase
  https://doi.org/10.1016/j.commatsci.2022.111670} {\bibfield  {journal}
  {\bibinfo  {journal} {Computational Materials Science}\ }\textbf {\bibinfo
  {volume} {213}},\ \bibinfo {pages} {111670} (\bibinfo {year}
  {2022})}\BibitemShut {NoStop}%
\bibitem [{\citenamefont {Wang}\ \emph {et~al.}(2022)\citenamefont {Wang},
  \citenamefont {Bedoya-Pinto}, \citenamefont {Blei}, \citenamefont {Dismukes},
  \citenamefont {Hamo}, \citenamefont {Jenkins}, \citenamefont {Koperski},
  \citenamefont {Liu}, \citenamefont {Sun}, \citenamefont {Telford},
  \citenamefont {Kim}, \citenamefont {Augustin}, \citenamefont {Vool},
  \citenamefont {Yin}, \citenamefont {Li}, \citenamefont {Falin}, \citenamefont
  {Dean}, \citenamefont {Casanova}, \citenamefont {Evans}, \citenamefont
  {Chshiev}, \citenamefont {Mishchenko}, \citenamefont {Petrovic},
  \citenamefont {He}, \citenamefont {Zhao}, \citenamefont {Tsen}, \citenamefont
  {Gerardot}, \citenamefont {Brotons-Gisbert}, \citenamefont {Guguchia},
  \citenamefont {Roy}, \citenamefont {Tongay}, \citenamefont {Wang},
  \citenamefont {Hasan}, \citenamefont {Wrachtrup}, \citenamefont {Yacoby},
  \citenamefont {Fert}, \citenamefont {Parkin}, \citenamefont {Novoselov},
  \citenamefont {Dai}, \citenamefont {Balicas},\ and\ \citenamefont
  {Santos}}]{Review2022-VanDerWaals}%
  \BibitemOpen
  \bibfield  {author} {\bibinfo {author} {\bibfnamefont {Q.~H.}\ \bibnamefont
  {Wang}}, \bibinfo {author} {\bibfnamefont {A.}~\bibnamefont {Bedoya-Pinto}},
  \bibinfo {author} {\bibfnamefont {M.}~\bibnamefont {Blei}}, \bibinfo {author}
  {\bibfnamefont {A.~H.}\ \bibnamefont {Dismukes}}, \bibinfo {author}
  {\bibfnamefont {A.}~\bibnamefont {Hamo}}, \bibinfo {author} {\bibfnamefont
  {S.}~\bibnamefont {Jenkins}}, \bibinfo {author} {\bibfnamefont
  {M.}~\bibnamefont {Koperski}}, \bibinfo {author} {\bibfnamefont
  {Y.}~\bibnamefont {Liu}}, \bibinfo {author} {\bibfnamefont {Q.}~\bibnamefont
  {Sun}}, \bibinfo {author} {\bibfnamefont {E.~J.}\ \bibnamefont {Telford}},
  \bibinfo {author} {\bibfnamefont {H.~H.}\ \bibnamefont {Kim}}, \bibinfo
  {author} {\bibfnamefont {M.}~\bibnamefont {Augustin}}, \bibinfo {author}
  {\bibfnamefont {U.}~\bibnamefont {Vool}}, \bibinfo {author} {\bibfnamefont
  {J.}~\bibnamefont {Yin}}, \bibinfo {author} {\bibfnamefont {L.~H.}\
  \bibnamefont {Li}}, \bibinfo {author} {\bibfnamefont {A.}~\bibnamefont
  {Falin}}, \bibinfo {author} {\bibfnamefont {C.~R.}\ \bibnamefont {Dean}},
  \bibinfo {author} {\bibfnamefont {F.}~\bibnamefont {Casanova}}, \bibinfo
  {author} {\bibfnamefont {R.~F.~L.}\ \bibnamefont {Evans}}, \bibinfo {author}
  {\bibfnamefont {M.}~\bibnamefont {Chshiev}}, \bibinfo {author} {\bibfnamefont
  {A.}~\bibnamefont {Mishchenko}}, \bibinfo {author} {\bibfnamefont
  {C.}~\bibnamefont {Petrovic}}, \bibinfo {author} {\bibfnamefont
  {R.}~\bibnamefont {He}}, \bibinfo {author} {\bibfnamefont {L.}~\bibnamefont
  {Zhao}}, \bibinfo {author} {\bibfnamefont {A.~W.}\ \bibnamefont {Tsen}},
  \bibinfo {author} {\bibfnamefont {B.~D.}\ \bibnamefont {Gerardot}}, \bibinfo
  {author} {\bibfnamefont {M.}~\bibnamefont {Brotons-Gisbert}}, \bibinfo
  {author} {\bibfnamefont {Z.}~\bibnamefont {Guguchia}}, \bibinfo {author}
  {\bibfnamefont {X.}~\bibnamefont {Roy}}, \bibinfo {author} {\bibfnamefont
  {S.}~\bibnamefont {Tongay}}, \bibinfo {author} {\bibfnamefont
  {Z.}~\bibnamefont {Wang}}, \bibinfo {author} {\bibfnamefont {M.~Z.}\
  \bibnamefont {Hasan}}, \bibinfo {author} {\bibfnamefont {J.}~\bibnamefont
  {Wrachtrup}}, \bibinfo {author} {\bibfnamefont {A.}~\bibnamefont {Yacoby}},
  \bibinfo {author} {\bibfnamefont {A.}~\bibnamefont {Fert}}, \bibinfo {author}
  {\bibfnamefont {S.}~\bibnamefont {Parkin}}, \bibinfo {author} {\bibfnamefont
  {K.~S.}\ \bibnamefont {Novoselov}}, \bibinfo {author} {\bibfnamefont
  {P.}~\bibnamefont {Dai}}, \bibinfo {author} {\bibfnamefont {L.}~\bibnamefont
  {Balicas}}, \ and\ \bibinfo {author} {\bibfnamefont {E.~J.~G.}\ \bibnamefont
  {Santos}},\ }\href {\doibase 10.1021/acsnano.1c09150} {\bibfield  {journal}
  {\bibinfo  {journal} {ACS Nano}\ }\textbf {\bibinfo {volume} {16}},\ \bibinfo
  {pages} {6960} (\bibinfo {year} {2022})},\ \bibinfo {note} {pMID: 35442017},\
  \Eprint {http://arxiv.org/abs/https://doi.org/10.1021/acsnano.1c09150}
  {https://doi.org/10.1021/acsnano.1c09150} \BibitemShut {NoStop}%
\bibitem [{\citenamefont {Lee}\ \emph {et~al.}(2016)\citenamefont {Lee},
  \citenamefont {Lee}, \citenamefont {Ryoo}, \citenamefont {Kang},
  \citenamefont {Kim}, \citenamefont {Kim}, \citenamefont {Park}, \citenamefont
  {Park},\ and\ \citenamefont {Cheong}}]{FePS3-2016}%
  \BibitemOpen
  \bibfield  {author} {\bibinfo {author} {\bibfnamefont {J.}~\bibnamefont
  {Lee}}, \bibinfo {author} {\bibfnamefont {S.}~\bibnamefont {Lee}}, \bibinfo
  {author} {\bibfnamefont {J.~H.}\ \bibnamefont {Ryoo}}, \bibinfo {author}
  {\bibfnamefont {S.}~\bibnamefont {Kang}}, \bibinfo {author} {\bibfnamefont
  {T.~Y.}\ \bibnamefont {Kim}}, \bibinfo {author} {\bibfnamefont
  {P.}~\bibnamefont {Kim}}, \bibinfo {author} {\bibfnamefont {C.}~\bibnamefont
  {Park}}, \bibinfo {author} {\bibfnamefont {J.}~\bibnamefont {Park}}, \ and\
  \bibinfo {author} {\bibfnamefont {H.}~\bibnamefont {Cheong}},\ }\href
  {\doibase 10.1021/acs.nanolett.6b03052} {\bibfield  {journal} {\bibinfo
  {journal} {Nano Letters}\ }\textbf {\bibinfo {volume} {16}},\ \bibinfo
  {pages} {7433} (\bibinfo {year} {2016})},\ \bibinfo {note} {pMID: 27960508},\
  \Eprint {http://arxiv.org/abs/https://doi.org/10.1021/acs.nanolett.6b03052}
  {https://doi.org/10.1021/acs.nanolett.6b03052} \BibitemShut {NoStop}%
\bibitem [{\citenamefont {Mermin}\ and\ \citenamefont
  {Wagner}(1966)}]{PhysRevLett.17.1133}%
  \BibitemOpen
  \bibfield  {author} {\bibinfo {author} {\bibfnamefont {N.~D.}\ \bibnamefont
  {Mermin}}\ and\ \bibinfo {author} {\bibfnamefont {H.}~\bibnamefont
  {Wagner}},\ }\href {\doibase 10.1103/PhysRevLett.17.1133} {\bibfield
  {journal} {\bibinfo  {journal} {Phys. Rev. Lett.}\ }\textbf {\bibinfo
  {volume} {17}},\ \bibinfo {pages} {1133} (\bibinfo {year}
  {1966})}\BibitemShut {NoStop}%
\bibitem [{\citenamefont {Hovan\ifmmode~\check{c}\else \v{c}\fi{}\'{\i}k}\
  \emph {et~al.}(2023)\citenamefont {Hovan\ifmmode~\check{c}\else
  \v{c}\fi{}\'{\i}k}, \citenamefont {Kratochv\'{\i}lov\'a}, \citenamefont
  {Haidamak}, \citenamefont {Dole\ifmmode~\check{z}\else \v{z}\fi{}al},
  \citenamefont {Carva}, \citenamefont {Bendov\'a}, \citenamefont
  {Prokle\ifmmode~\check{s}\else \v{s}\fi{}ka}, \citenamefont {Proschek},
  \citenamefont {M\'{\i}\ifmmode~\check{s}\else \v{s}\fi{}ek}, \citenamefont
  {Gorbunov}, \citenamefont {Kotek}, \citenamefont {Sechovsk\'y},\ and\
  \citenamefont {Posp\'{\i}\ifmmode~\check{s}\else \v{s}\fi{}il}}]{VBr3-2023}%
  \BibitemOpen
  \bibfield  {author} {\bibinfo {author} {\bibfnamefont {D.}~\bibnamefont
  {Hovan\ifmmode~\check{c}\else \v{c}\fi{}\'{\i}k}}, \bibinfo {author}
  {\bibfnamefont {M.}~\bibnamefont {Kratochv\'{\i}lov\'a}}, \bibinfo {author}
  {\bibfnamefont {T.}~\bibnamefont {Haidamak}}, \bibinfo {author}
  {\bibfnamefont {P.}~\bibnamefont {Dole\ifmmode~\check{z}\else \v{z}\fi{}al}},
  \bibinfo {author} {\bibfnamefont {K.}~\bibnamefont {Carva}}, \bibinfo
  {author} {\bibfnamefont {A.}~\bibnamefont {Bendov\'a}}, \bibinfo {author}
  {\bibfnamefont {J.}~\bibnamefont {Prokle\ifmmode~\check{s}\else
  \v{s}\fi{}ka}}, \bibinfo {author} {\bibfnamefont {P.}~\bibnamefont
  {Proschek}}, \bibinfo {author} {\bibfnamefont {M.}~\bibnamefont
  {M\'{\i}\ifmmode~\check{s}\else \v{s}\fi{}ek}}, \bibinfo {author}
  {\bibfnamefont {D.~I.}\ \bibnamefont {Gorbunov}}, \bibinfo {author}
  {\bibfnamefont {J.}~\bibnamefont {Kotek}}, \bibinfo {author} {\bibfnamefont
  {V.}~\bibnamefont {Sechovsk\'y}}, \ and\ \bibinfo {author} {\bibfnamefont
  {J.}~\bibnamefont {Posp\'{\i}\ifmmode~\check{s}\else \v{s}\fi{}il}},\ }\href
  {\doibase 10.1103/PhysRevB.108.104416} {\bibfield  {journal} {\bibinfo
  {journal} {Phys. Rev. B}\ }\textbf {\bibinfo {volume} {108}},\ \bibinfo
  {pages} {104416} (\bibinfo {year} {2023})}\BibitemShut {NoStop}%
\bibitem [{\citenamefont {Huang}\ \emph {et~al.}(2017)\citenamefont {Huang},
  \citenamefont {Clark}, \citenamefont {Navarro-Moratalla}, \citenamefont
  {Klein}, \citenamefont {Cheng}, \citenamefont {Seyler}, \citenamefont
  {Zhong}, \citenamefont {Schmidgall}, \citenamefont {McGuire}, \citenamefont
  {Cobden} \emph {et~al.}}]{huang2017layer}%
  \BibitemOpen
  \bibfield  {author} {\bibinfo {author} {\bibfnamefont {B.}~\bibnamefont
  {Huang}}, \bibinfo {author} {\bibfnamefont {G.}~\bibnamefont {Clark}},
  \bibinfo {author} {\bibfnamefont {E.}~\bibnamefont {Navarro-Moratalla}},
  \bibinfo {author} {\bibfnamefont {D.~R.}\ \bibnamefont {Klein}}, \bibinfo
  {author} {\bibfnamefont {R.}~\bibnamefont {Cheng}}, \bibinfo {author}
  {\bibfnamefont {K.~L.}\ \bibnamefont {Seyler}}, \bibinfo {author}
  {\bibfnamefont {D.}~\bibnamefont {Zhong}}, \bibinfo {author} {\bibfnamefont
  {E.}~\bibnamefont {Schmidgall}}, \bibinfo {author} {\bibfnamefont {M.~A.}\
  \bibnamefont {McGuire}}, \bibinfo {author} {\bibfnamefont {D.~H.}\
  \bibnamefont {Cobden}},  \emph {et~al.},\ }\href@noop {} {\bibfield
  {journal} {\bibinfo  {journal} {Nature}\ }\textbf {\bibinfo {volume} {546}},\
  \bibinfo {pages} {270} (\bibinfo {year} {2017})}\BibitemShut {NoStop}%
\bibitem [{\citenamefont {Chen}\ \emph {et~al.}(2024)\citenamefont {Chen},
  \citenamefont {Teng}, \citenamefont {Hu}, \citenamefont {Ye}, \citenamefont
  {Granroth}, \citenamefont {Yi}, \citenamefont {Chung}, \citenamefont
  {Birgeneau},\ and\ \citenamefont {Dai}}]{chen2024thermal}%
  \BibitemOpen
  \bibfield  {author} {\bibinfo {author} {\bibfnamefont {L.}~\bibnamefont
  {Chen}}, \bibinfo {author} {\bibfnamefont {X.}~\bibnamefont {Teng}}, \bibinfo
  {author} {\bibfnamefont {D.}~\bibnamefont {Hu}}, \bibinfo {author}
  {\bibfnamefont {F.}~\bibnamefont {Ye}}, \bibinfo {author} {\bibfnamefont
  {G.~E.}\ \bibnamefont {Granroth}}, \bibinfo {author} {\bibfnamefont
  {M.}~\bibnamefont {Yi}}, \bibinfo {author} {\bibfnamefont {J.-H.}\
  \bibnamefont {Chung}}, \bibinfo {author} {\bibfnamefont {R.~J.}\ \bibnamefont
  {Birgeneau}}, \ and\ \bibinfo {author} {\bibfnamefont {P.}~\bibnamefont
  {Dai}},\ }\href@noop {} {\bibfield  {journal} {\bibinfo  {journal} {npj
  Quantum Materials}\ }\textbf {\bibinfo {volume} {9}},\ \bibinfo {pages} {40}
  (\bibinfo {year} {2024})}\BibitemShut {NoStop}%
\bibitem [{\citenamefont {Wildes}\ \emph {et~al.}(2020)\citenamefont {Wildes},
  \citenamefont {Lan\ifmmode~\mbox{\c{c}}\else \c{c}\fi{}on}, \citenamefont
  {Chan}, \citenamefont {Weickert}, \citenamefont {Harrison}, \citenamefont
  {Simonet}, \citenamefont {Zhitomirsky}, \citenamefont {Gvozdikova},
  \citenamefont {Ziman},\ and\ \citenamefont {R\o{}nnow}}]{WildesPRB-2020}%
  \BibitemOpen
  \bibfield  {author} {\bibinfo {author} {\bibfnamefont {A.~R.}\ \bibnamefont
  {Wildes}}, \bibinfo {author} {\bibfnamefont {D.}~\bibnamefont
  {Lan\ifmmode~\mbox{\c{c}}\else \c{c}\fi{}on}}, \bibinfo {author}
  {\bibfnamefont {M.~K.}\ \bibnamefont {Chan}}, \bibinfo {author}
  {\bibfnamefont {F.}~\bibnamefont {Weickert}}, \bibinfo {author}
  {\bibfnamefont {N.}~\bibnamefont {Harrison}}, \bibinfo {author}
  {\bibfnamefont {V.}~\bibnamefont {Simonet}}, \bibinfo {author} {\bibfnamefont
  {M.~E.}\ \bibnamefont {Zhitomirsky}}, \bibinfo {author} {\bibfnamefont
  {M.~V.}\ \bibnamefont {Gvozdikova}}, \bibinfo {author} {\bibfnamefont
  {T.}~\bibnamefont {Ziman}}, \ and\ \bibinfo {author} {\bibfnamefont {H.~M.}\
  \bibnamefont {R\o{}nnow}},\ }\href {\doibase 10.1103/PhysRevB.101.024415}
  {\bibfield  {journal} {\bibinfo  {journal} {Phys. Rev. B}\ }\textbf {\bibinfo
  {volume} {101}},\ \bibinfo {pages} {024415} (\bibinfo {year}
  {2020})}\BibitemShut {NoStop}%
\bibitem [{\citenamefont {Bob\'{a}k}\ \emph {et~al.}(2016)\citenamefont
  {Bob\'{a}k}, \citenamefont {Lu\v{c}ivjansk\'{y}}, \citenamefont
  {\v{Z}ukovi\v{c}}, \citenamefont {Borovsk\'{y}},\ and\ \citenamefont
  {Balcerzak}}]{BOBAK20162693}%
  \BibitemOpen
  \bibfield  {author} {\bibinfo {author} {\bibfnamefont {A.}~\bibnamefont
  {Bob\'{a}k}}, \bibinfo {author} {\bibfnamefont {T.}~\bibnamefont
  {Lu\v{c}ivjansk\'{y}}}, \bibinfo {author} {\bibfnamefont {M.}~\bibnamefont
  {\v{Z}ukovi\v{c}}}, \bibinfo {author} {\bibfnamefont {M.}~\bibnamefont
  {Borovsk\'{y}}}, \ and\ \bibinfo {author} {\bibfnamefont {T.}~\bibnamefont
  {Balcerzak}},\ }\href {\doibase
  https://doi.org/10.1016/j.physleta.2016.06.019} {\bibfield  {journal}
  {\bibinfo  {journal} {Physics Letters A}\ }\textbf {\bibinfo {volume}
  {380}},\ \bibinfo {pages} {2693} (\bibinfo {year} {2016})}\BibitemShut
  {NoStop}%
\bibitem [{\citenamefont {Schmidt}\ and\ \citenamefont
  {Godoy}(2021)}]{SCHMIDT2021168151}%
  \BibitemOpen
  \bibfield  {author} {\bibinfo {author} {\bibfnamefont {M.}~\bibnamefont
  {Schmidt}}\ and\ \bibinfo {author} {\bibfnamefont {P.}~\bibnamefont
  {Godoy}},\ }\href {\doibase https://doi.org/10.1016/j.jmmm.2021.168151}
  {\bibfield  {journal} {\bibinfo  {journal} {Journal of Magnetism and Magnetic
  Materials}\ }\textbf {\bibinfo {volume} {537}},\ \bibinfo {pages} {168151}
  (\bibinfo {year} {2021})}\BibitemShut {NoStop}%
\bibitem [{\citenamefont {Batista}\ \emph {et~al.}(2025)\citenamefont
  {Batista}, \citenamefont {Schmidt},\ and\ \citenamefont
  {Zimmer}}]{batista2025correlatedclustermeanfieldapproach}%
  \BibitemOpen
  \bibfield  {author} {\bibinfo {author} {\bibfnamefont {C.~H.~D.}\
  \bibnamefont {Batista}}, \bibinfo {author} {\bibfnamefont {M.}~\bibnamefont
  {Schmidt}}, \ and\ \bibinfo {author} {\bibfnamefont {F.~M.}\ \bibnamefont
  {Zimmer}},\ }\href {https://arxiv.org/abs/2509.21512} {\bibfield  {journal}
  {\bibinfo  {journal} {arXiv preprint, arXiv:2509.21512}\ } (\bibinfo {year}
  {2025})},\ \Eprint {http://arxiv.org/abs/2509.21512} {arXiv:2509.21512
  [cond-mat.stat-mech]} \BibitemShut {NoStop}%
\bibitem [{\citenamefont {Gessert}\ \emph {et~al.}(2024)\citenamefont
  {Gessert}, \citenamefont {Weigel},\ and\ \citenamefont
  {Janke}}]{Entropy_2024}%
  \BibitemOpen
  \bibfield  {author} {\bibinfo {author} {\bibfnamefont {D.}~\bibnamefont
  {Gessert}}, \bibinfo {author} {\bibfnamefont {M.}~\bibnamefont {Weigel}}, \
  and\ \bibinfo {author} {\bibfnamefont {W.}~\bibnamefont {Janke}},\ }\href
  {\doibase 10.3390/e26110919} {\bibfield  {journal} {\bibinfo  {journal}
  {Entropy}\ }\textbf {\bibinfo {volume} {26}} (\bibinfo {year} {2024}),\
  10.3390/e26110919}\BibitemShut {NoStop}%
\bibitem [{\citenamefont {\v{Z}ukovi\v{c}}(2021)}]{Zukovic-2021}%
  \BibitemOpen
  \bibfield  {author} {\bibinfo {author} {\bibfnamefont {M.}~\bibnamefont
  {\v{Z}ukovi\v{c}}},\ }\href {\doibase
  https://doi.org/10.1016/j.physleta.2021.127405} {\bibfield  {journal}
  {\bibinfo  {journal} {Physics Letters A}\ }\textbf {\bibinfo {volume}
  {404}},\ \bibinfo {pages} {127405} (\bibinfo {year} {2021})}\BibitemShut
  {NoStop}%
\bibitem [{\citenamefont {Gessert}\ \emph {et~al.}(2025)\citenamefont
  {Gessert}, \citenamefont {Weigel},\ and\ \citenamefont
  {Janke}}]{gessert2025}%
  \BibitemOpen
  \bibfield  {author} {\bibinfo {author} {\bibfnamefont {D.}~\bibnamefont
  {Gessert}}, \bibinfo {author} {\bibfnamefont {M.}~\bibnamefont {Weigel}}, \
  and\ \bibinfo {author} {\bibfnamefont {W.}~\bibnamefont {Janke}},\ }\href
  {https://arxiv.org/abs/2509.03414} {\bibfield  {journal} {\bibinfo  {journal}
  {arXiv preprint, arXiv:2509.03414}\ } (\bibinfo {year} {2025})},\ \Eprint
  {http://arxiv.org/abs/2509.03414} {arXiv:2509.03414 [cond-mat.stat-mech]}
  \BibitemShut {NoStop}%
\bibitem [{\citenamefont {Azhari}\ \emph {et~al.}(2025)\citenamefont {Azhari},
  \citenamefont {Jang},\ and\ \citenamefont {Yu}}]{AZHARI2025108412}%
  \BibitemOpen
  \bibfield  {author} {\bibinfo {author} {\bibfnamefont {M.}~\bibnamefont
  {Azhari}}, \bibinfo {author} {\bibfnamefont {H.}~\bibnamefont {Jang}}, \ and\
  \bibinfo {author} {\bibfnamefont {U.}~\bibnamefont {Yu}},\ }\href {\doibase
  https://doi.org/10.1016/j.rinp.2025.108412} {\bibfield  {journal} {\bibinfo
  {journal} {Results in Physics}\ }\textbf {\bibinfo {volume} {76}},\ \bibinfo
  {pages} {108412} (\bibinfo {year} {2025})}\BibitemShut {NoStop}%
\bibitem [{\citenamefont {Li}\ \emph {et~al.}(2025)\citenamefont {Li},
  \citenamefont {Tseng},\ and\ \citenamefont {Jiang}}]{li2025}%
  \BibitemOpen
  \bibfield  {author} {\bibinfo {author} {\bibfnamefont {S.-W.}\ \bibnamefont
  {Li}}, \bibinfo {author} {\bibfnamefont {Y.-H.}\ \bibnamefont {Tseng}}, \
  and\ \bibinfo {author} {\bibfnamefont {F.-J.}\ \bibnamefont {Jiang}},\ }\href
  {https://arxiv.org/abs/2509.11437} {\bibfield  {journal} {\bibinfo  {journal}
  {arXiv preprint, arXiv:2509.11437}\ } (\bibinfo {year} {2025})},\ \Eprint
  {http://arxiv.org/abs/2509.11437} {arXiv:2509.11437 [hep-lat]} \BibitemShut
  {NoStop}%
\bibitem [{\citenamefont {Dias}\ and\ \citenamefont
  {Schmidt}(2023)}]{Pietro2023-PRB}%
  \BibitemOpen
  \bibfield  {author} {\bibinfo {author} {\bibfnamefont {P.~F.}\ \bibnamefont
  {Dias}}\ and\ \bibinfo {author} {\bibfnamefont {M.}~\bibnamefont {Schmidt}},\
  }\href {\doibase 10.1103/PhysRevB.108.014436} {\bibfield  {journal} {\bibinfo
   {journal} {Phys. Rev. B}\ }\textbf {\bibinfo {volume} {108}},\ \bibinfo
  {pages} {014436} (\bibinfo {year} {2023})}\BibitemShut {NoStop}%
\bibitem [{\citenamefont {Dias}\ \emph {et~al.}(2024)\citenamefont {Dias},
  \citenamefont {Krindges}, \citenamefont {Morais}, \citenamefont {Zimmer},
  \citenamefont {Mohylna}, \citenamefont {\v{Z}ukovi\v{c}},\ and\ \citenamefont
  {Schmidt}}]{Pietro2024-MC}%
  \BibitemOpen
  \bibfield  {author} {\bibinfo {author} {\bibfnamefont {P.}~\bibnamefont
  {Dias}}, \bibinfo {author} {\bibfnamefont {A.}~\bibnamefont {Krindges}},
  \bibinfo {author} {\bibfnamefont {C.}~\bibnamefont {Morais}}, \bibinfo
  {author} {\bibfnamefont {F.}~\bibnamefont {Zimmer}}, \bibinfo {author}
  {\bibfnamefont {M.}~\bibnamefont {Mohylna}}, \bibinfo {author} {\bibfnamefont
  {M.}~\bibnamefont {\v{Z}ukovi\v{c}}}, \ and\ \bibinfo {author} {\bibfnamefont
  {M.}~\bibnamefont {Schmidt}},\ }\href {\doibase
  https://doi.org/10.1016/j.jmmm.2024.172282} {\bibfield  {journal} {\bibinfo
  {journal} {Journal of Magnetism and Magnetic Materials}\ }\textbf {\bibinfo
  {volume} {604}},\ \bibinfo {pages} {172282} (\bibinfo {year}
  {2024})}\BibitemShut {NoStop}%
\bibitem [{\citenamefont {Yamamoto}\ \emph {et~al.}(2015)\citenamefont
  {Yamamoto}, \citenamefont {Marmorini},\ and\ \citenamefont
  {Danshita}}]{PhysRevLett.114.027201}%
  \BibitemOpen
  \bibfield  {author} {\bibinfo {author} {\bibfnamefont {D.}~\bibnamefont
  {Yamamoto}}, \bibinfo {author} {\bibfnamefont {G.}~\bibnamefont {Marmorini}},
  \ and\ \bibinfo {author} {\bibfnamefont {I.}~\bibnamefont {Danshita}},\
  }\href {\doibase 10.1103/PhysRevLett.114.027201} {\bibfield  {journal}
  {\bibinfo  {journal} {Phys. Rev. Lett.}\ }\textbf {\bibinfo {volume} {114}},\
  \bibinfo {pages} {027201} (\bibinfo {year} {2015})}\BibitemShut {NoStop}%
\bibitem [{\citenamefont {Roos}\ and\ \citenamefont
  {Schmidt}(2024)}]{Roos2024}%
  \BibitemOpen
  \bibfield  {author} {\bibinfo {author} {\bibfnamefont {M.}~\bibnamefont
  {Roos}}\ and\ \bibinfo {author} {\bibfnamefont {M.}~\bibnamefont {Schmidt}},\
  }\href {\doibase https://doi.org/10.1016/j.physa.2024.129979} {\bibfield
  {journal} {\bibinfo  {journal} {Physica A: Statistical Mechanics and its
  Applications}\ }\textbf {\bibinfo {volume} {651}},\ \bibinfo {pages} {129979}
  (\bibinfo {year} {2024})}\BibitemShut {NoStop}%
\bibitem [{\citenamefont {Yamamoto}\ \emph {et~al.}(2020)\citenamefont
  {Yamamoto}, \citenamefont {Suzuki}, \citenamefont {Marmorini}, \citenamefont
  {Okazaki},\ and\ \citenamefont {Furukawa}}]{PhysRevLett.125.057204}%
  \BibitemOpen
  \bibfield  {author} {\bibinfo {author} {\bibfnamefont {D.}~\bibnamefont
  {Yamamoto}}, \bibinfo {author} {\bibfnamefont {C.}~\bibnamefont {Suzuki}},
  \bibinfo {author} {\bibfnamefont {G.}~\bibnamefont {Marmorini}}, \bibinfo
  {author} {\bibfnamefont {S.}~\bibnamefont {Okazaki}}, \ and\ \bibinfo
  {author} {\bibfnamefont {N.}~\bibnamefont {Furukawa}},\ }\href {\doibase
  10.1103/PhysRevLett.125.057204} {\bibfield  {journal} {\bibinfo  {journal}
  {Phys. Rev. Lett.}\ }\textbf {\bibinfo {volume} {125}},\ \bibinfo {pages}
  {057204} (\bibinfo {year} {2020})}\BibitemShut {NoStop}%
\bibitem [{\citenamefont {Rossato}\ \emph {et~al.}(2023)\citenamefont
  {Rossato}, \citenamefont {Zimmer}, \citenamefont {Morais},\ and\
  \citenamefont {Schmidt}}]{Rossato2023}%
  \BibitemOpen
  \bibfield  {author} {\bibinfo {author} {\bibfnamefont {L.~C.}\ \bibnamefont
  {Rossato}}, \bibinfo {author} {\bibfnamefont {F.}~\bibnamefont {Zimmer}},
  \bibinfo {author} {\bibfnamefont {C.}~\bibnamefont {Morais}}, \ and\ \bibinfo
  {author} {\bibfnamefont {M.}~\bibnamefont {Schmidt}},\ }\href {\doibase
  https://doi.org/10.1016/j.physa.2023.128778} {\bibfield  {journal} {\bibinfo
  {journal} {Physica A: Statistical Mechanics and its Applications}\ }\textbf
  {\bibinfo {volume} {621}},\ \bibinfo {pages} {128778} (\bibinfo {year}
  {2023})}\BibitemShut {NoStop}%
\bibitem [{\citenamefont {Singhania}\ and\ \citenamefont
  {Kumar}(2020)}]{PhysRevB.101.064403}%
  \BibitemOpen
  \bibfield  {author} {\bibinfo {author} {\bibfnamefont {A.}~\bibnamefont
  {Singhania}}\ and\ \bibinfo {author} {\bibfnamefont {S.}~\bibnamefont
  {Kumar}},\ }\href {\doibase 10.1103/PhysRevB.101.064403} {\bibfield
  {journal} {\bibinfo  {journal} {Phys. Rev. B}\ }\textbf {\bibinfo {volume}
  {101}},\ \bibinfo {pages} {064403} (\bibinfo {year} {2020})}\BibitemShut
  {NoStop}%
\bibitem [{\citenamefont {Roos}\ \emph {et~al.}(2024)\citenamefont {Roos},
  \citenamefont {Muhl}, \citenamefont {Schmidt}, \citenamefont {Morais},\ and\
  \citenamefont {Zimmer}}]{PhysRevE.109.014144}%
  \BibitemOpen
  \bibfield  {author} {\bibinfo {author} {\bibfnamefont {M.}~\bibnamefont
  {Roos}}, \bibinfo {author} {\bibfnamefont {I.~F.}\ \bibnamefont {Muhl}},
  \bibinfo {author} {\bibfnamefont {M.}~\bibnamefont {Schmidt}}, \bibinfo
  {author} {\bibfnamefont {C.~V.}\ \bibnamefont {Morais}}, \ and\ \bibinfo
  {author} {\bibfnamefont {F.~M.}\ \bibnamefont {Zimmer}},\ }\href {\doibase
  10.1103/PhysRevE.109.014144} {\bibfield  {journal} {\bibinfo  {journal}
  {Phys. Rev. E}\ }\textbf {\bibinfo {volume} {109}},\ \bibinfo {pages}
  {014144} (\bibinfo {year} {2024})}\BibitemShut {NoStop}%
\bibitem [{\citenamefont {Kud{\={o}}}\ and\ \citenamefont
  {Katsura}(1976)}]{PTP_1976}%
  \BibitemOpen
  \bibfield  {author} {\bibinfo {author} {\bibfnamefont {T.}~\bibnamefont
  {Kud{\={o}}}}\ and\ \bibinfo {author} {\bibfnamefont {S.}~\bibnamefont
  {Katsura}},\ }\href {\doibase 10.1143/PTP.56.435} {\bibfield  {journal}
  {\bibinfo  {journal} {Progress of Theoretical Physics}\ }\textbf {\bibinfo
  {volume} {56}},\ \bibinfo {pages} {435} (\bibinfo {year} {1976})}\BibitemShut
  {NoStop}%
\bibitem [{\citenamefont {Kerrai}\ \emph {et~al.}(2024)\citenamefont {Kerrai},
  \citenamefont {Jalal}, \citenamefont {Zaim}, \citenamefont {Hasnaoui},
  \citenamefont {Kerouad},\ and\ \citenamefont
  {El~Bouziani}}]{kerrai2024study}%
  \BibitemOpen
  \bibfield  {author} {\bibinfo {author} {\bibfnamefont {H.}~\bibnamefont
  {Kerrai}}, \bibinfo {author} {\bibfnamefont {E.}~\bibnamefont {Jalal}},
  \bibinfo {author} {\bibfnamefont {A.}~\bibnamefont {Zaim}}, \bibinfo {author}
  {\bibfnamefont {A.}~\bibnamefont {Hasnaoui}}, \bibinfo {author}
  {\bibfnamefont {M.}~\bibnamefont {Kerouad}}, \ and\ \bibinfo {author}
  {\bibfnamefont {M.}~\bibnamefont {El~Bouziani}},\ }\href@noop {} {\bibfield
  {journal} {\bibinfo  {journal} {Applied Physics A}\ }\textbf {\bibinfo
  {volume} {130}},\ \bibinfo {pages} {816} (\bibinfo {year}
  {2024})}\BibitemShut {NoStop}%
\bibitem [{\citenamefont {Griffiths}(1970)}]{PhysRevLett.24.715}%
  \BibitemOpen
  \bibfield  {author} {\bibinfo {author} {\bibfnamefont {R.~B.}\ \bibnamefont
  {Griffiths}},\ }\href {\doibase 10.1103/PhysRevLett.24.715} {\bibfield
  {journal} {\bibinfo  {journal} {Phys. Rev. Lett.}\ }\textbf {\bibinfo
  {volume} {24}},\ \bibinfo {pages} {715} (\bibinfo {year} {1970})}\BibitemShut
  {NoStop}%
\bibitem [{\citenamefont {Stishov}\ and\ \citenamefont
  {Petrova}(2020)}]{Stishov2020}%
  \BibitemOpen
  \bibfield  {author} {\bibinfo {author} {\bibfnamefont {S.~M.}\ \bibnamefont
  {Stishov}}\ and\ \bibinfo {author} {\bibfnamefont {A.~E.}\ \bibnamefont
  {Petrova}},\ }\href {\doibase 10.1134/S106377612011014X} {\bibfield
  {journal} {\bibinfo  {journal} {Journal of Experimental and Theoretical
  Physics}\ }\textbf {\bibinfo {volume} {131}},\ \bibinfo {pages} {1056}
  (\bibinfo {year} {2020})}\BibitemShut {NoStop}%
\bibitem [{\citenamefont {Moueddene}\ \emph
  {et~al.}(2024{\natexlab{a}})\citenamefont {Moueddene}, \citenamefont {Fytas},
  \citenamefont {Holovatch}, \citenamefont {Kenna},\ and\ \citenamefont
  {Berche}}]{Moueddene_2024}%
  \BibitemOpen
  \bibfield  {author} {\bibinfo {author} {\bibfnamefont {L.}~\bibnamefont
  {Moueddene}}, \bibinfo {author} {\bibfnamefont {N.~G.}\ \bibnamefont
  {Fytas}}, \bibinfo {author} {\bibfnamefont {Y.}~\bibnamefont {Holovatch}},
  \bibinfo {author} {\bibfnamefont {R.}~\bibnamefont {Kenna}}, \ and\ \bibinfo
  {author} {\bibfnamefont {B.}~\bibnamefont {Berche}},\ }\href {\doibase
  10.1088/1742-5468/ad1d60} {\bibfield  {journal} {\bibinfo  {journal} {Journal
  of Statistical Mechanics: Theory and Experiment}\ }\textbf {\bibinfo {volume}
  {2024}},\ \bibinfo {pages} {023206} (\bibinfo {year}
  {2024}{\natexlab{a}})}\BibitemShut {NoStop}%
\bibitem [{\citenamefont {Moueddene}\ \emph
  {et~al.}(2024{\natexlab{b}})\citenamefont {Moueddene}, \citenamefont
  {Fytas},\ and\ \citenamefont {Berche}}]{PhysRevE.110.064144}%
  \BibitemOpen
  \bibfield  {author} {\bibinfo {author} {\bibfnamefont {L.}~\bibnamefont
  {Moueddene}}, \bibinfo {author} {\bibfnamefont {N.~G.}\ \bibnamefont
  {Fytas}}, \ and\ \bibinfo {author} {\bibfnamefont {B.}~\bibnamefont
  {Berche}},\ }\href {\doibase 10.1103/PhysRevE.110.064144} {\bibfield
  {journal} {\bibinfo  {journal} {Phys. Rev. E}\ }\textbf {\bibinfo {volume}
  {110}},\ \bibinfo {pages} {064144} (\bibinfo {year}
  {2024}{\natexlab{b}})}\BibitemShut {NoStop}%
\bibitem [{\citenamefont {Mataragkas}\ \emph
  {et~al.}(2025{\natexlab{a}})\citenamefont {Mataragkas}, \citenamefont
  {Vasilopoulos}, \citenamefont {Fytas},\ and\ \citenamefont
  {Kim}}]{PhysRevResearch.7.013214}%
  \BibitemOpen
  \bibfield  {author} {\bibinfo {author} {\bibfnamefont {D.}~\bibnamefont
  {Mataragkas}}, \bibinfo {author} {\bibfnamefont {A.}~\bibnamefont
  {Vasilopoulos}}, \bibinfo {author} {\bibfnamefont {N.~G.}\ \bibnamefont
  {Fytas}}, \ and\ \bibinfo {author} {\bibfnamefont {D.-H.}\ \bibnamefont
  {Kim}},\ }\href {\doibase 10.1103/PhysRevResearch.7.013214} {\bibfield
  {journal} {\bibinfo  {journal} {Phys. Rev. Res.}\ }\textbf {\bibinfo {volume}
  {7}},\ \bibinfo {pages} {013214} (\bibinfo {year}
  {2025}{\natexlab{a}})}\BibitemShut {NoStop}%
\bibitem [{\citenamefont {Mataragkas}\ \emph
  {et~al.}(2025{\natexlab{b}})\citenamefont {Mataragkas}, \citenamefont
  {Vasilopoulos}, \citenamefont {Fytas},\ and\ \citenamefont
  {Kim}}]{PhysRevResearch.7.033240}%
  \BibitemOpen
  \bibfield  {author} {\bibinfo {author} {\bibfnamefont {D.}~\bibnamefont
  {Mataragkas}}, \bibinfo {author} {\bibfnamefont {A.}~\bibnamefont
  {Vasilopoulos}}, \bibinfo {author} {\bibfnamefont {N.~G.}\ \bibnamefont
  {Fytas}}, \ and\ \bibinfo {author} {\bibfnamefont {D.-H.}\ \bibnamefont
  {Kim}},\ }\href {\doibase 10.1103/jfl3-f4kd} {\bibfield  {journal} {\bibinfo
  {journal} {Phys. Rev. Res.}\ }\textbf {\bibinfo {volume} {7}},\ \bibinfo
  {pages} {033240} (\bibinfo {year} {2025}{\natexlab{b}})}\BibitemShut
  {NoStop}%
\bibitem [{\citenamefont {Rocha-Neto}\ \emph {et~al.}(2023)\citenamefont
  {Rocha-Neto}, \citenamefont {Camelo-Neto}, \citenamefont {Nogueira},\ and\
  \citenamefont {Coutinho}}]{ROCHANETO2023129145}%
  \BibitemOpen
  \bibfield  {author} {\bibinfo {author} {\bibfnamefont {M.~J.}\ \bibnamefont
  {Rocha-Neto}}, \bibinfo {author} {\bibfnamefont {G.}~\bibnamefont
  {Camelo-Neto}}, \bibinfo {author} {\bibfnamefont {E.}~\bibnamefont
  {Nogueira}}, \ and\ \bibinfo {author} {\bibfnamefont {S.}~\bibnamefont
  {Coutinho}},\ }\href {\doibase https://doi.org/10.1016/j.physa.2023.129145}
  {\bibfield  {journal} {\bibinfo  {journal} {Physica A: Statistical Mechanics
  and its Applications}\ }\textbf {\bibinfo {volume} {629}},\ \bibinfo {pages}
  {129145} (\bibinfo {year} {2023})}\BibitemShut {NoStop}%
\bibitem [{\citenamefont {Saadi}\ \emph {et~al.}(2024)\citenamefont {Saadi},
  \citenamefont {Kerrai}, \citenamefont {Jalal}, \citenamefont {Elhani},
  \citenamefont {Al-Rajhi},\ and\ \citenamefont
  {Bouziani}}]{saadi2024critical}%
  \BibitemOpen
  \bibfield  {author} {\bibinfo {author} {\bibfnamefont {H.}~\bibnamefont
  {Saadi}}, \bibinfo {author} {\bibfnamefont {H.}~\bibnamefont {Kerrai}},
  \bibinfo {author} {\bibfnamefont {E.}~\bibnamefont {Jalal}}, \bibinfo
  {author} {\bibfnamefont {A.}~\bibnamefont {Elhani}}, \bibinfo {author}
  {\bibfnamefont {A.}~\bibnamefont {Al-Rajhi}}, \ and\ \bibinfo {author}
  {\bibfnamefont {M.~E.}\ \bibnamefont {Bouziani}},\ }\href@noop {} {\bibfield
  {journal} {\bibinfo  {journal} {Applied Physics A}\ }\textbf {\bibinfo
  {volume} {130}},\ \bibinfo {pages} {918} (\bibinfo {year}
  {2024})}\BibitemShut {NoStop}%
\bibitem [{\citenamefont {Dom\'{\i}nguez}\ \emph {et~al.}(2021)\citenamefont
  {Dom\'{\i}nguez}, \citenamefont {Lopetegui},\ and\ \citenamefont
  {Mulet}}]{Dominguez2021-CV}%
  \BibitemOpen
  \bibfield  {author} {\bibinfo {author} {\bibfnamefont {E.}~\bibnamefont
  {Dom\'{\i}nguez}}, \bibinfo {author} {\bibfnamefont {C.~E.}\ \bibnamefont
  {Lopetegui}}, \ and\ \bibinfo {author} {\bibfnamefont {R.}~\bibnamefont
  {Mulet}},\ }\href {\doibase 10.1103/PhysRevB.104.014205} {\bibfield
  {journal} {\bibinfo  {journal} {Phys. Rev. B}\ }\textbf {\bibinfo {volume}
  {104}},\ \bibinfo {pages} {014205} (\bibinfo {year} {2021})}\BibitemShut
  {NoStop}%
\bibitem [{\citenamefont {Hu}\ and\ \citenamefont
  {Charbonneau}(2021{\natexlab{a}})}]{Hu2021-TM}%
  \BibitemOpen
  \bibfield  {author} {\bibinfo {author} {\bibfnamefont {Y.}~\bibnamefont
  {Hu}}\ and\ \bibinfo {author} {\bibfnamefont {P.}~\bibnamefont
  {Charbonneau}},\ }\href {\doibase 10.1103/PhysRevB.104.144429} {\bibfield
  {journal} {\bibinfo  {journal} {Phys. Rev. B}\ }\textbf {\bibinfo {volume}
  {104}},\ \bibinfo {pages} {144429} (\bibinfo {year}
  {2021}{\natexlab{a}})}\BibitemShut {NoStop}%
\bibitem [{\citenamefont {Chatelain}(2025)}]{Chatelain2025-TN}%
  \BibitemOpen
  \bibfield  {author} {\bibinfo {author} {\bibfnamefont {C.}~\bibnamefont
  {Chatelain}},\ }\href {\doibase 10.1103/PhysRevE.111.024109} {\bibfield
  {journal} {\bibinfo  {journal} {Phys. Rev. E}\ }\textbf {\bibinfo {volume}
  {111}},\ \bibinfo {pages} {024109} (\bibinfo {year} {2025})}\BibitemShut
  {NoStop}%
\bibitem [{\citenamefont {Roos}\ \emph {et~al.}(2026)\citenamefont {Roos},
  \citenamefont {Morais}, \citenamefont {Zimmer},\ and\ \citenamefont
  {Schmidt}}]{ROOS2026131093}%
  \BibitemOpen
  \bibfield  {author} {\bibinfo {author} {\bibfnamefont {M.}~\bibnamefont
  {Roos}}, \bibinfo {author} {\bibfnamefont {C.}~\bibnamefont {Morais}},
  \bibinfo {author} {\bibfnamefont {F.}~\bibnamefont {Zimmer}}, \ and\ \bibinfo
  {author} {\bibfnamefont {M.}~\bibnamefont {Schmidt}},\ }\href {\doibase
  https://doi.org/10.1016/j.physa.2025.131093} {\bibfield  {journal} {\bibinfo
  {journal} {Physica A: Statistical Mechanics and its Applications}\ }\textbf
  {\bibinfo {volume} {681}},\ \bibinfo {pages} {131093} (\bibinfo {year}
  {2026})}\BibitemShut {NoStop}%
\bibitem [{\citenamefont {Park}(2016)}]{Park2016-VanDerWaals}%
  \BibitemOpen
  \bibfield  {author} {\bibinfo {author} {\bibfnamefont {J.}~\bibnamefont
  {Park}},\ }\href {\doibase 10.1088/0953-8984/28/30/301001} {\bibfield
  {journal} {\bibinfo  {journal} {Journal of Physics: Condensed Matter}\
  }\textbf {\bibinfo {volume} {28}},\ \bibinfo {pages} {301001} (\bibinfo
  {year} {2016})}\BibitemShut {NoStop}%
\bibitem [{\citenamefont {G\'omez~Albarrac\'{\i}n}\ \emph
  {et~al.}(2018)\citenamefont {G\'omez~Albarrac\'{\i}n}, \citenamefont
  {Rosales},\ and\ \citenamefont {Serra}}]{Albarracin2018}%
  \BibitemOpen
  \bibfield  {author} {\bibinfo {author} {\bibfnamefont {F.~A.}\ \bibnamefont
  {G\'omez~Albarrac\'{\i}n}}, \bibinfo {author} {\bibfnamefont {H.~D.}\
  \bibnamefont {Rosales}}, \ and\ \bibinfo {author} {\bibfnamefont
  {P.}~\bibnamefont {Serra}},\ }\href {\doibase 10.1103/PhysRevE.98.012139}
  {\bibfield  {journal} {\bibinfo  {journal} {Phys. Rev. E}\ }\textbf {\bibinfo
  {volume} {98}},\ \bibinfo {pages} {012139} (\bibinfo {year}
  {2018})}\BibitemShut {NoStop}%
\bibitem [{\citenamefont {Bergman}\ \emph {et~al.}(2007)\citenamefont
  {Bergman}, \citenamefont {Alicea}, \citenamefont {Gull}, \citenamefont
  {Trebst},\ and\ \citenamefont {Balents}}]{bergman2007order}%
  \BibitemOpen
  \bibfield  {author} {\bibinfo {author} {\bibfnamefont {D.}~\bibnamefont
  {Bergman}}, \bibinfo {author} {\bibfnamefont {J.}~\bibnamefont {Alicea}},
  \bibinfo {author} {\bibfnamefont {E.}~\bibnamefont {Gull}}, \bibinfo {author}
  {\bibfnamefont {S.}~\bibnamefont {Trebst}}, \ and\ \bibinfo {author}
  {\bibfnamefont {L.}~\bibnamefont {Balents}},\ }\href@noop {} {\bibfield
  {journal} {\bibinfo  {journal} {Nature Physics}\ }\textbf {\bibinfo {volume}
  {3}},\ \bibinfo {pages} {487} (\bibinfo {year} {2007})}\BibitemShut {NoStop}%
\bibitem [{\citenamefont {{Villain, J.}}\ \emph {et~al.}(1980)\citenamefont
  {{Villain, J.}}, \citenamefont {{Bidaux, R.}}, \citenamefont {{Carton,
  J.-P.}},\ and\ \citenamefont {{Conte, R.}}}]{OBD_1980}%
  \BibitemOpen
  \bibfield  {author} {\bibinfo {author} {\bibnamefont {{Villain, J.}}},
  \bibinfo {author} {\bibnamefont {{Bidaux, R.}}}, \bibinfo {author}
  {\bibnamefont {{Carton, J.-P.}}}, \ and\ \bibinfo {author} {\bibnamefont
  {{Conte, R.}}},\ }\href {\doibase 10.1051/jphys:0198000410110126300}
  {\bibfield  {journal} {\bibinfo  {journal} {J. Phys. France}\ }\textbf
  {\bibinfo {volume} {41}},\ \bibinfo {pages} {1263} (\bibinfo {year}
  {1980})}\BibitemShut {NoStop}%
\bibitem [{\citenamefont {Fisher}\ and\ \citenamefont
  {Nelson}(1974)}]{PhysRevLett.32.1350}%
  \BibitemOpen
  \bibfield  {author} {\bibinfo {author} {\bibfnamefont {M.~E.}\ \bibnamefont
  {Fisher}}\ and\ \bibinfo {author} {\bibfnamefont {D.~R.}\ \bibnamefont
  {Nelson}},\ }\href {\doibase 10.1103/PhysRevLett.32.1350} {\bibfield
  {journal} {\bibinfo  {journal} {Phys. Rev. Lett.}\ }\textbf {\bibinfo
  {volume} {32}},\ \bibinfo {pages} {1350} (\bibinfo {year}
  {1974})}\BibitemShut {NoStop}%
\bibitem [{\citenamefont {Zaim}\ \emph {et~al.}(2024)\citenamefont {Zaim},
  \citenamefont {Kerrai}, \citenamefont {Zaim}, \citenamefont {Kerouad},\ and\
  \citenamefont {Zaim}}]{ZAIM2024113536}%
  \BibitemOpen
  \bibfield  {author} {\bibinfo {author} {\bibfnamefont {N.}~\bibnamefont
  {Zaim}}, \bibinfo {author} {\bibfnamefont {H.}~\bibnamefont {Kerrai}},
  \bibinfo {author} {\bibfnamefont {M.}~\bibnamefont {Zaim}}, \bibinfo {author}
  {\bibfnamefont {M.}~\bibnamefont {Kerouad}}, \ and\ \bibinfo {author}
  {\bibfnamefont {A.}~\bibnamefont {Zaim}},\ }\href {\doibase
  https://doi.org/10.1016/j.vacuum.2024.113536} {\bibfield  {journal} {\bibinfo
   {journal} {Vacuum}\ }\textbf {\bibinfo {volume} {228}},\ \bibinfo {pages}
  {113536} (\bibinfo {year} {2024})}\BibitemShut {NoStop}%
\bibitem [{\citenamefont {Kerrai}\ \emph {et~al.}(2025)\citenamefont {Kerrai},
  \citenamefont {Zaim},\ and\ \citenamefont {Kerouad}}]{PSS_2025}%
  \BibitemOpen
  \bibfield  {author} {\bibinfo {author} {\bibfnamefont {H.}~\bibnamefont
  {Kerrai}}, \bibinfo {author} {\bibfnamefont {A.}~\bibnamefont {Zaim}}, \ and\
  \bibinfo {author} {\bibfnamefont {M.}~\bibnamefont {Kerouad}},\ }\href
  {\doibase https://doi.org/10.1002/pssr.202400330} {\bibfield  {journal}
  {\bibinfo  {journal} {physica status solidi (RRL) – Rapid Research
  Letters}\ }\textbf {\bibinfo {volume} {19}},\ \bibinfo {pages} {2400330}
  (\bibinfo {year} {2025})}\BibitemShut {NoStop}%
\bibitem [{\citenamefont {Guerrero}\ \emph {et~al.}(2015)\citenamefont
  {Guerrero}, \citenamefont {Stariolo},\ and\ \citenamefont
  {Almarza}}]{VariacionalPRE-2015}%
  \BibitemOpen
  \bibfield  {author} {\bibinfo {author} {\bibfnamefont {A.~I.}\ \bibnamefont
  {Guerrero}}, \bibinfo {author} {\bibfnamefont {D.~A.}\ \bibnamefont
  {Stariolo}}, \ and\ \bibinfo {author} {\bibfnamefont {N.~G.}\ \bibnamefont
  {Almarza}},\ }\href {\doibase 10.1103/PhysRevE.91.052123} {\bibfield
  {journal} {\bibinfo  {journal} {Phys. Rev. E}\ }\textbf {\bibinfo {volume}
  {91}},\ \bibinfo {pages} {052123} (\bibinfo {year} {2015})}\BibitemShut
  {NoStop}%
\bibitem [{\citenamefont {Acevedo}\ \emph {et~al.}(2021)\citenamefont
  {Acevedo}, \citenamefont {Lamas},\ and\ \citenamefont
  {Pujol}}]{EfetivoPRB-2021}%
  \BibitemOpen
  \bibfield  {author} {\bibinfo {author} {\bibfnamefont {S.}~\bibnamefont
  {Acevedo}}, \bibinfo {author} {\bibfnamefont {C.~A.}\ \bibnamefont {Lamas}},
  \ and\ \bibinfo {author} {\bibfnamefont {P.}~\bibnamefont {Pujol}},\ }\href
  {\doibase 10.1103/PhysRevB.104.214412} {\bibfield  {journal} {\bibinfo
  {journal} {Phys. Rev. B}\ }\textbf {\bibinfo {volume} {104}},\ \bibinfo
  {pages} {214412} (\bibinfo {year} {2021})}\BibitemShut {NoStop}%
\bibitem [{\citenamefont {Vanhecke}\ \emph {et~al.}(2021)\citenamefont
  {Vanhecke}, \citenamefont {Colbois}, \citenamefont {Vanderstraeten},
  \citenamefont {Verstraete},\ and\ \citenamefont {Mila}}]{TensorPRR-2021}%
  \BibitemOpen
  \bibfield  {author} {\bibinfo {author} {\bibfnamefont {B.}~\bibnamefont
  {Vanhecke}}, \bibinfo {author} {\bibfnamefont {J.}~\bibnamefont {Colbois}},
  \bibinfo {author} {\bibfnamefont {L.}~\bibnamefont {Vanderstraeten}},
  \bibinfo {author} {\bibfnamefont {F.}~\bibnamefont {Verstraete}}, \ and\
  \bibinfo {author} {\bibfnamefont {F.}~\bibnamefont {Mila}},\ }\href {\doibase
  10.1103/PhysRevResearch.3.013041} {\bibfield  {journal} {\bibinfo  {journal}
  {Phys. Rev. Res.}\ }\textbf {\bibinfo {volume} {3}},\ \bibinfo {pages}
  {013041} (\bibinfo {year} {2021})}\BibitemShut {NoStop}%
\bibitem [{\citenamefont {Hu}\ and\ \citenamefont
  {Charbonneau}(2021{\natexlab{b}})}]{MatrizPRB-2021}%
  \BibitemOpen
  \bibfield  {author} {\bibinfo {author} {\bibfnamefont {Y.}~\bibnamefont
  {Hu}}\ and\ \bibinfo {author} {\bibfnamefont {P.}~\bibnamefont
  {Charbonneau}},\ }\href {\doibase 10.1103/PhysRevB.104.144429} {\bibfield
  {journal} {\bibinfo  {journal} {Phys. Rev. B}\ }\textbf {\bibinfo {volume}
  {104}},\ \bibinfo {pages} {144429} (\bibinfo {year}
  {2021}{\natexlab{b}})}\BibitemShut {NoStop}%
\bibitem [{\citenamefont {Le~Mardel\'e}\ \emph {et~al.}(2024)\citenamefont
  {Le~Mardel\'e}, \citenamefont {El~Mendili}, \citenamefont {Zhitomirsky},
  \citenamefont {Mohelsky}, \citenamefont {Jana}, \citenamefont {Plutnarova},
  \citenamefont {Sofer}, \citenamefont {Faugeras}, \citenamefont {Potemski},\
  and\ \citenamefont {Orlita}}]{PhysRevB.109.134410}%
  \BibitemOpen
  \bibfield  {author} {\bibinfo {author} {\bibfnamefont {F.}~\bibnamefont
  {Le~Mardel\'e}}, \bibinfo {author} {\bibfnamefont {A.}~\bibnamefont
  {El~Mendili}}, \bibinfo {author} {\bibfnamefont {M.~E.}\ \bibnamefont
  {Zhitomirsky}}, \bibinfo {author} {\bibfnamefont {I.}~\bibnamefont
  {Mohelsky}}, \bibinfo {author} {\bibfnamefont {D.}~\bibnamefont {Jana}},
  \bibinfo {author} {\bibfnamefont {I.}~\bibnamefont {Plutnarova}}, \bibinfo
  {author} {\bibfnamefont {Z.}~\bibnamefont {Sofer}}, \bibinfo {author}
  {\bibfnamefont {C.}~\bibnamefont {Faugeras}}, \bibinfo {author}
  {\bibfnamefont {M.}~\bibnamefont {Potemski}}, \ and\ \bibinfo {author}
  {\bibfnamefont {M.}~\bibnamefont {Orlita}},\ }\href {\doibase
  10.1103/PhysRevB.109.134410} {\bibfield  {journal} {\bibinfo  {journal}
  {Phys. Rev. B}\ }\textbf {\bibinfo {volume} {109}},\ \bibinfo {pages}
  {134410} (\bibinfo {year} {2024})}\BibitemShut {NoStop}%
\end{thebibliography}%

\end{document}